\documentclass[12pt]{article}

\usepackage{authblk}
\usepackage{amsmath}
\usepackage{graphicx}
\usepackage{enumerate}
\usepackage{natbib}
\usepackage{url}
\usepackage{tcolorbox}

\usepackage{enumitem}
\usepackage{amsmath}    
\usepackage{amssymb}    
\usepackage{amsthm}     

\theoremstyle{plain}
\newtheorem{theorem}{Theorem}

\theoremstyle{definition}
\newtheorem{definition}[theorem]{Definition}

\usepackage{caption}
\usepackage{graphicx}
\usepackage{xcolor}
\usepackage{algorithm}
\usepackage{algorithmic}
\usepackage{datetime2} 

\usepackage{tikz}
\usetikzlibrary{positioning,arrows.meta}
\usepackage{booktabs}  

\begin{document}

\setlength{\parskip}{0.4\baselineskip plus 2pt}
\setlength{\parindent}{0pt}

\def\spacingset#1{\renewcommand{\baselinestretch}%
{#1}\small\normalsize} \spacingset{1}

\title{\bf From Global to Local Correlation: Geometric Decomposition of Statistical Inference\thanks{Research reported in this publication was supported by grant number INV-048956 from the Gates Foundation.}}
\author{Pawel Gajer$^{1}$\thanks{Corresponding author: pgajer@gmail.com}\; and Jacques Ravel$^{1}$}
\date{%
\small
$^1$University of Maryland School of Medicine
}
\maketitle

{\renewcommand{\thefootnote}{}\footnotetext{Software: The methods are implemented in the R package \texttt{gflow}, available at \url{https://github.com/pgajer/gflow}.}}

\begin{abstract}
  Understanding feature-outcome associations in high-dimensional data remains
  challenging when relationships vary across subpopulations, yet standard
  methods assuming global associations miss context-dependent patterns, reducing
  statistical power and interpretability. We develop a geometric decomposition
  framework offering two strategies for partitioning inference problems into
  regional analyses on data-derived Riemannian graphs. Gradient flow
  decomposition uses path-monotonicity-validated discrete Morse theory to
  partition samples into gradient flow cells where outcomes exhibit monotonic
  behavior. Co-monotonicity decomposition utilizes vertex-level coefficients
  that provide context-dependent versions of the classical Pearson correlation:
  these coefficients measure edge-based directional concordance between outcome
  and features, or between feature pairs, defining embeddings of samples into
  association space. These embeddings induce Riemannian k-NN graphs on which
  biclustering identifies co-monotonicity cells (coherent regions) and feature
  modules. This extends naturally to multi-modal integration across multiple
  feature sets. Both strategies apply independently or jointly, with Bayesian
  posterior sampling providing credible intervals.
\end{abstract}


\newpage
\section*{1. Introduction}

Understanding which features associate with outcomes in high-dimensional data
remains one of the central challenges in modern statistical analysis. Consider a
microbiome study investigating spontaneous preterm birth
\cite{fettweis2019vaginal,elovitz2019cervicovaginal}, where researchers measure
hundreds of bacterial taxa across hundreds of samples. A particular phylotype
might promote disease risk in women with one microbial community composition
while showing no effect or even protective association in women with different
community structures. Traditional correlation and regression methods assume
global, homogeneous relationships across all samples, averaging over such
conflicting signals and potentially concluding that no association exists when
in fact multiple distinct context-dependent mechanisms operate simultaneously.

This phenomenon of spatially heterogeneous associations appears throughout
high-dimensional biomarker studies. In single-cell genomics, gene expression
programs exhibit cell-type-specific relationships with outcomes
\cite{trapnell2015defining}. In spatial transcriptomics, tissue architecture
creates regions where the same molecular features play different functional
roles \cite{berglund2018spatial}. In ecological data, species interactions vary
across environmental gradients \cite{liu2022environmental}. The common thread is
that the ambient high-dimensional feature space contains subpopulations or
regions where statistical relationships differ fundamentally, yet these regions
are unknown a priori and must be discovered from data. When associations of
opposite sign cancel in global analyses, investigators miss biologically
meaningful mechanisms entirely, and the resulting models provide no basis for
stratified interventions or personalized predictions.

The Pearson correlation coefficient
\citep{pearson1895notes,pearson1896regression}, one of the foundational measures
in theoretical statistics, exemplifies both the power and the limitation of
global association measures. Its normalization structure, dividing covariance by
the product of standard deviations, ensures interpretable coefficients bounded
in $[-1,1]$ regardless of measurement scales. This elegant formulation has made
correlation ubiquitous in statistical practice. However, computing associations
through deviations from global means treats all observations as equally related,
discarding any spatial or relational structure among samples. Our contribution
develops co-monotonicity coefficients as geometric refinements of Pearson
correlation: we preserve the proven normalization structure while replacing
global mean deviations with local edge-based directional concordance, yielding
vertex-level measures that respect the geometric organization of
high-dimensional data and reveal how associations vary across regions.

What is needed to realize this vision is a framework that naturally discovers
the contextual structure latent in high-dimensional data and partitions the
sample space into regions of homogeneous associative behavior. The partition
itself should emerge from geometric properties of the data, which encode the
complex interactions between features implicit in the ambient representation,
rather than being imposed through arbitrary choices. By deconvoluting these
hidden association patterns and making them explicit through geometric
decomposition, the framework enables rigorous uncertainty quantification that
respects both the spatial dependencies in graph-structured data and the
exploratory nature of discovering structure and testing associations
simultaneously.

A fundamental characteristic of high-dimensional data poses both challenge and
opportunity: while datasets may contain hundreds or thousands of measured
features, the underlying system often operates through far fewer degrees of
freedom. Genomic studies measure tens of thousands of genes, yet cellular states
often organize along a small number of developmental or functional axes
\cite{trapnell2014dynamics}. Microbiome samples with hundreds of bacterial taxa
concentrate near community type structures rather than filling the ambient space
\cite{ravel2011vaginal,arumugam2011enterotypes}. Social network data with
extensive demographic and behavioral variables frequently reduces to a handful
of latent factors \cite{hoff2002latent}. The manifold hypothesis
\cite{tenenbaum2000global,fefferman2016testing} formalizes this observation:
high-dimensional data typically concentrates near low-dimensional manifolds
embedded in the ambient feature space. While real data rarely forms smooth
manifolds (exhibiting noise, stratification, and singular structures) the core
insight remains valid that intrinsic dimensionality is far lower than nominal
dimensionality, and that this geometric structure can be exploited for
inference.

Geometric data analysis adopts a coordinate-free perspective, representing data
through graphs or simplicial complexes that capture intrinsic relationships
between samples. Rather than computing $E[Y|X]$ where $X \in \mathbb{R}^p$, we
shift to computing $E[Y|G(X,y)]$ where $G(X,y)$ denotes a density-aware graph
constructed from the predictor matrix $X$ and observed response $y$. This
reformulation makes the geometric structure explicit: statistical inference
operates directly on the graph that encodes how samples relate to one another.

We work specifically with Riemannian graphs that possess rich local metric
structure beyond simple edge weights. These graphs arise naturally from
intersection k-nearest neighbor constructions interpreted through nerve
complexes. Each vertex $v$ corresponds to a sample, and the vertex mass $m_0(v)$
reflects the local density of samples in that region. Edges connect nearby
samples, with edge masses $m_1(e)$ capturing the geometric extent of
neighborhoods that share the edge. The Riemannian structure on the graph is
encoded by the vertex mass matrix $M_0$ (diagonal, with entries $m_0(v)$) and
the edge inner product matrix $M_1$. While $M_0$ is determined by vertex masses
alone, $M_1$ encodes more complex geometric relationships: beyond edge masses,
it captures angular information between incident edges through inner products
computed from the symmetrized graph Laplacian, defining a complete inner product
structure on the space of edge chains \cite{gajer2025geometry}. The normalized
graph Laplacian $L_{\text{norm}} = M_0^{-1/2} L_0 M_0^{-1/2}$, constructed from
these masses, governs diffusion processes on the graph
\cite{coifman2006diffusion} and enables spectral filtering methods for signal
recovery \cite{smola2003kernels,belkin2006manifold}.

Our approach builds fundamentally on Morse-Smale regression, introduced by
Gerber et al. \cite{gerber2013morse}. Their pioneering work brought ideas from
Morse theory in differential topology
\cite{morse1934calculus,witten1982supersymmetry} into the statistical inference
context, using gradient flow analysis to partition the feature space into cells
determined by pairs of local extrema of the conditional expectation surface,
then performing separate regression analyses within each cell. The framework
elegantly connects differential topology with statistical practice, providing
interpretable regional decompositions that respect the geometry of the
prediction landscape. However, translating this continuous theory to the
discrete, noisy setting of finite sample data presents several fundamental
challenges. First, robust estimation of conditional expectations on graphs
constructed from finite samples remains difficult. Moreover, even if the
conditional expectation estimate were a smooth function, sampling it on a finite
point set inevitably introduces spurious local extrema that do not correspond to
genuine features of the underlying continuous function. Without methods to
distinguish signal from noise in the extrema structure, the resulting cell
decomposition may fragment the space unnecessarily or preserve spurious
features. Second, real data-derived graphs typically contain long edges that
connect distant points, creating basin jumping artifacts where gradient
trajectories incorrectly cross between basins by following a single long edge
rather than a path through intermediate samples.

Third, applying standard linear models within each gradient flow cell assumes
that local linearity captures associations once heterogeneity is resolved
through decomposition. However, even if relationships within cells are indeed
locally linear, high-dimensional feature sets typically exhibit strong
multicollinearity, with many features highly correlated with each other.
Lasso-type regularization \cite{tibshirani1996regression} can partially address
this through automated feature selection, but different lasso variants handle
correlated features in problematic ways: standard lasso tends to arbitrarily
select one representative from a group of correlated features, while elastic net
\cite{zou2005regularization} includes entire groups but with potentially
excessive shrinkage. Moreover, if investigators wish to model how features
interact with each other to influence outcomes, interaction terms must be
pre-specified and included in the regression. With hundreds or thousands of
features, the combinatorial explosion of potential interactions becomes
computationally intractable and statistically underpowered
\cite{bellman1961adaptive,fan2010selective}. These challenges of
multicollinearity, arbitrary feature selection, and unmodeled interactions
suggest that alternative approaches to quantifying associations within cells may
prove more robust than fitting linear models. Finally, the discrete-to-discrete
translation requires careful treatment of the graph structure itself: working
with mutual k-nearest neighbor graphs with simple edge weights may not capture
the richer geometric structure that density-aware Riemannian metrics provide.

\begin{center}
\begin{minipage}[t]{.48\linewidth}
\includegraphics[width=\linewidth]{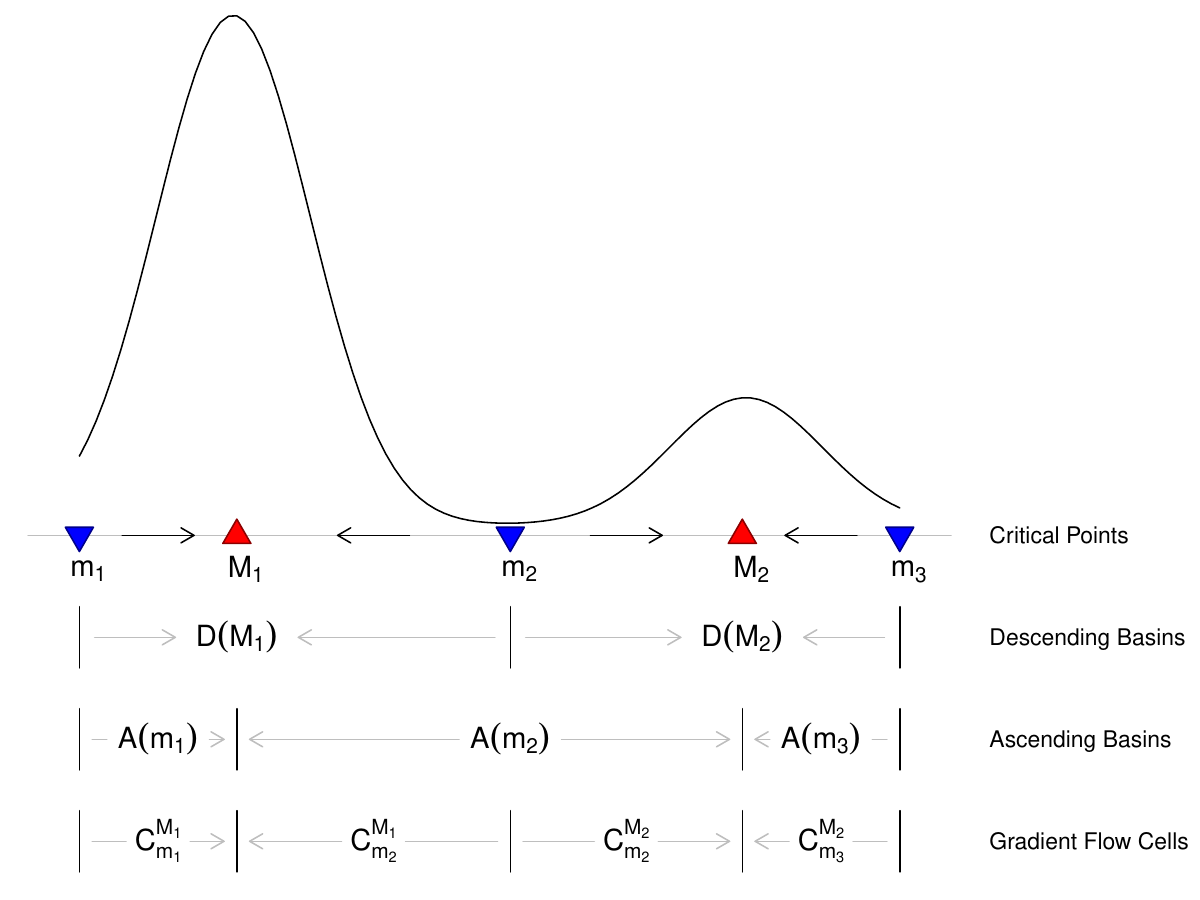}
\end{minipage}\hfill
\begin{minipage}[t]{.48\linewidth}
\includegraphics[width=\linewidth]{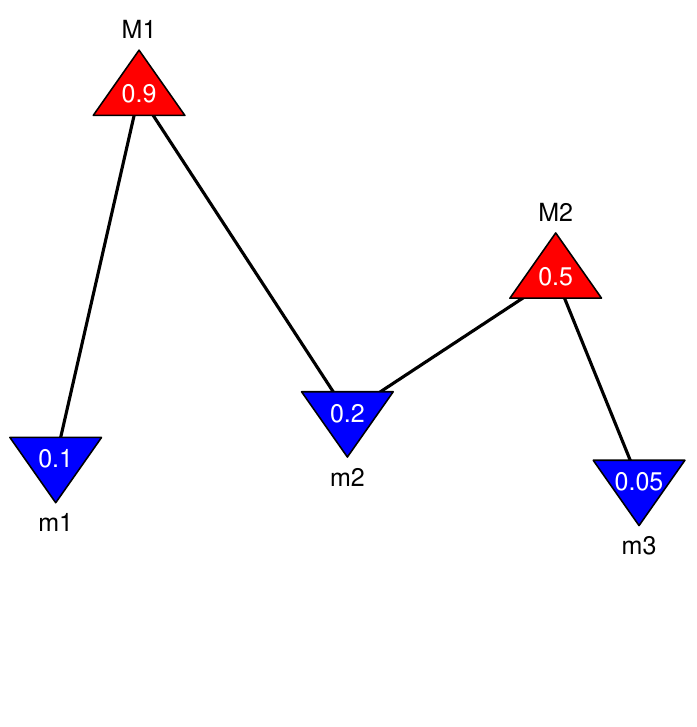}
\end{minipage}
\captionof{figure}{Domain decomposition via gradient flow. Panel A:
One-dimensional function with local minima ($m_1$, $m_2$, $m_3$) and maxima
($M_1$, $M_2$), showing gradient directions and partitioning into
ascending/descending basins and gradient flow cells. Panel B: Gradient flow
graph with vertices as critical points (labeled with function values) and edges
connecting minimum-maximum pairs, enabling monotonic statistical modeling within
each cell.}
\label{intro:fig1}
\end{center}

Our approach addresses these challenges while extending the Morse-Smale
framework in two fundamental directions. Unlike variable selection or dimension
reduction approaches that retain all samples while modifying the feature
representation
\cite{tibshirani1996regression,fan2010selective,tenenbaum2000global,roweis2000nonlinear},
geometric decomposition stratifies the sample space itself into regions where
different features may be relevant or where the same features operate through
different mechanisms.

\begin{center}
\includegraphics[width=\textwidth]{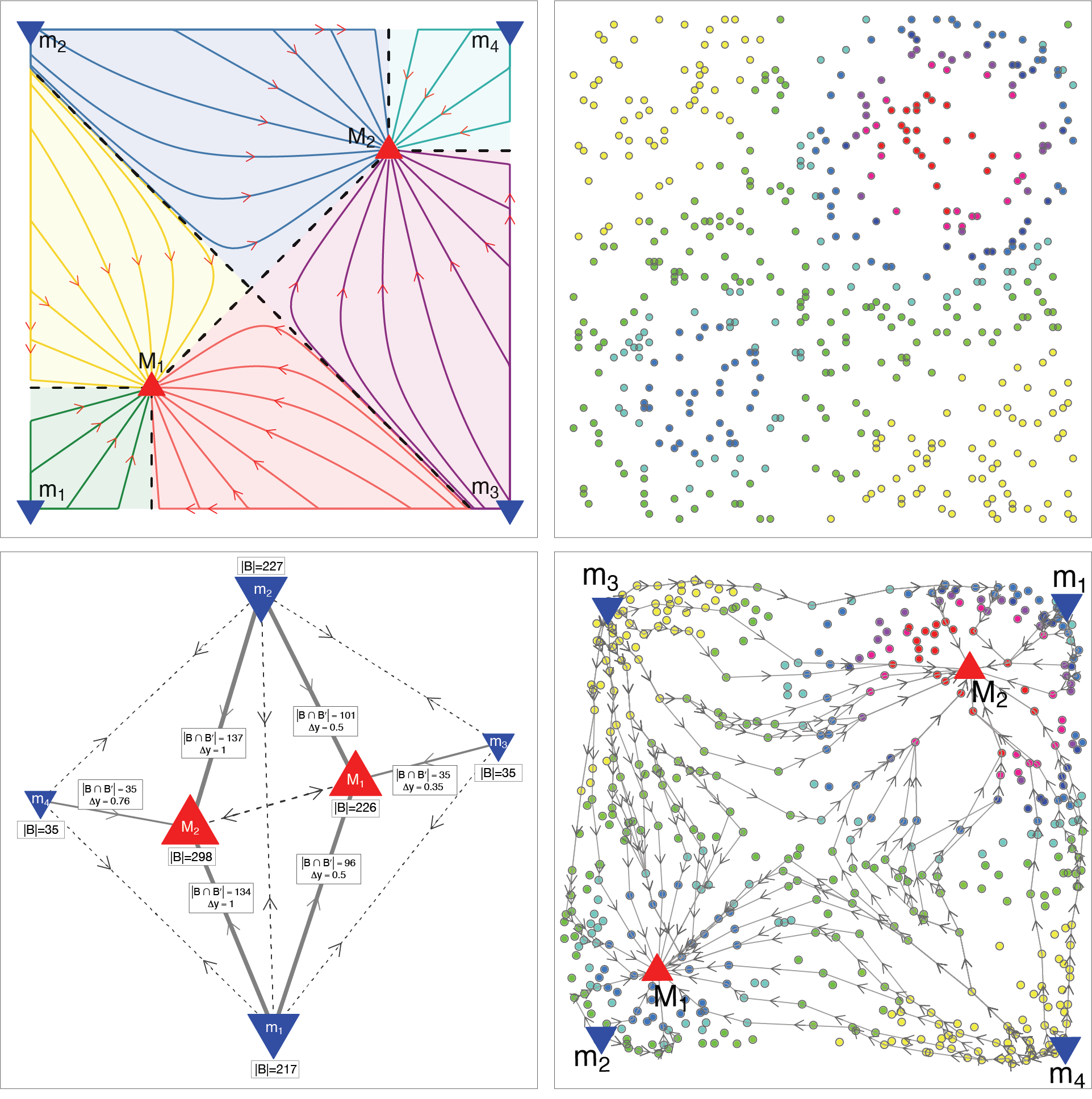}
\captionof{figure}{Gradient flow decomposition of a two-Gaussian mixture on
$[0,1]^2$. Top-left: Continuous function with critical points, gradient
trajectories, and cell boundaries. Top-right: Random sample with color-coded
function values. Bottom-right: k-nearest neighbor graph (k = 36) with selected
gradient trajectories. Bottom-left: Gradient flow complex where solid edges
connect minimum-maximum pairs and dashed edges represent minimum-minimum cells
with saddle points; edges shown for cells with at least 25 points.}
\label{intro:fig2}
\end{center}
Figures \ref{intro:fig1} and \ref{intro:fig2} illustrate these concepts.
Figure \ref{intro:fig1} demonstrates gradient-flow decomposition in one
dimension, showing how a function with multiple local extrema naturally
partitions into regions with homogeneous statistical behavior. Critical points
(local minima and maxima) form boundaries defining ascending and descending
basins, whose intersections create gradient-flow cells with monotonic behavior.

The gradient flow graph serves as a one-dimensional skeleton capturing the
essential shape of the prediction landscape, with edges representing natural
non-linear generalizations of principal directions in classical statistical
methods. Figure \ref{intro:fig2} extends this framework to two dimensions,
revealing additional complexities: non-trivial cell boundaries with irregular
geometries, meaningful transition regions through saddle points, and the
challenge of distinguishing statistically significant features from spurious
patterns arising from finite sampling.

The first strategy employs gradient flow analysis to partition samples based on
the geometry of the outcome surface. Given a smoothed estimate
$\hat{y}: V \to \mathbb{R}$ of the conditional expectation on graph vertices, we
construct the discrete gradient flow by iteratively moving from each vertex
toward neighbors with increasing (for ascending flow) or decreasing (for
descending flow) function values. The trajectories of these flows partition
vertices into basins of attraction: all vertices whose ascending flows terminate
at the same local minimum belong to that minimum's ascending basin, and
similarly for descending flows toward local maxima. The intersections of
ascending and descending basins define gradient flow cells
$C(m, M) = B(m) \cap B(M)$ where the outcome function exhibits monotonic
behavior. This construction realizes a discrete Morse-Smale complex on the
graph, partitioning the sample space into regions where the outcome surface has
uniform qualitative structure. We address the long edge problem through path
monotonicity validation: for edges connecting distant points, we verify that the
outcome function remains monotonic along the shortest alternative path. Only
edges passing this validation participate in gradient flow computation,
preventing basin jumping artifacts.

The second strategy emerged from considering how to quantify feature
associations within gradient flow cells. Since the smoothed outcome $\hat{y}$
exhibits monotonic behavior along gradient trajectories within each cell, a
natural approach to identifying associated features would measure their
monotonicity along these same trajectories. This path-based perspective,
treating monotonicity as a functor of paths and individual functions, suggested
an immediate generalization: rather than measuring single-function monotonicity
along paths, we could measure co-monotonicity between pairs of functions,
quantifying whether they vary together or oppositely along edges. A further
abstraction yields vertex-level measures by aggregating edge-wise
co-monotonicity over neighborhoods, producing coefficients that capture
directional concordance locally while being independent of any particular path
structure.

Co-monotonicity coefficients measure directional concordance between functions
at individual graph vertices: for functions $y, z: V \to \mathbb{R}$, the
coefficient at vertex $v$ quantifies whether $y$ and $z$ tend to change together
(positive co-monotonicity), opposite (negative), or independently (near zero)
across edges incident to $v$. By computing co-monotonicity between an outcome
$y$ and features in a set $Z = \{z_1, \ldots, z_m\}$, as well as between feature
pairs, we obtain association profiles for each sample. Figure~\ref{intro:fig3}
illustrates these profiles in a microbiome application, where hierarchical
clustering reveals coherent blocks of samples with similar association patterns
and features with coordinated outcome relationships. These profiles define
embeddings into association space, on which we construct k-nearest neighbor
graphs and apply spectral biclustering
\cite{dhillon2001co-clustering,kluger2003spectral} to identify co-monotonicity
cells—vertex regions and feature modules exhibiting coherent multivariate
association patterns. Where gradient flow partitions based on how outcomes vary
spatially, co-monotonicity partitioning discovers regions based on which
features associate with outcomes and how features relate to each other. The
framework extends naturally to multi-modal data integration, computing
association matrices across feature sets and their cross-associations to
construct embeddings that integrate information across modalities.

\begin{center}
\includegraphics[scale=0.5275]{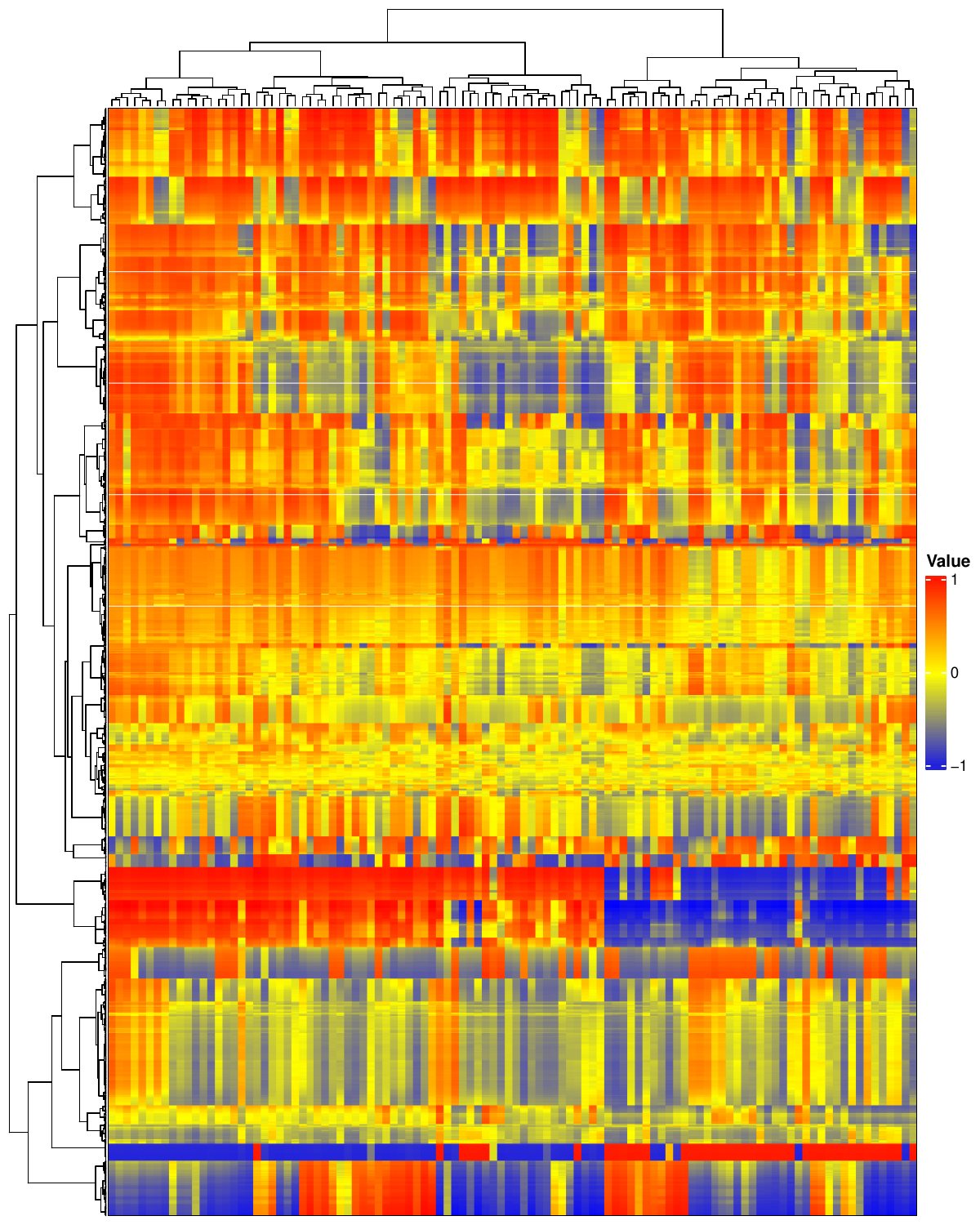}
\captionof{figure}{Co-monotonicity association profiles in vaginal microbiome
  data. Heatmap shows vertex-level smoothed (see Section 4.5) co-monotonicity
  coefficients between spontaneous preterm birth outcome and bacterial phylotype
  abundances across samples from pregnant women (rows: samples, columns:
  phylotypes). Hierarchical clustering on both axes reveals coherent blocks:
  samples (rows) group by shared association patterns, while phylotypes
  (columns) cluster by co-varying relationships with the outcome. Red indicates
  positive co-monotonicity (phylotype and outcome increase together), blue
  indicates negative co-monotonicity (inverse relationship), and yellow
  indicates independence. The block structure demonstrates how biclustering on
  these association profiles identifies co-monotonicity cells—regions where
  specific feature modules exhibit consistent outcome associations.}
\label{intro:fig3}
\end{center}

The two strategies can be applied independently or jointly. Independent
application suits exploratory analysis discovering structure from one
perspective. Joint application enables assessment of concordance: do
outcome-driven gradient flow cells align with association-driven co-monotonicity
cells? High concordance suggests that regions of similar outcome values arise
from consistent underlying mechanisms, while discordance indicates that similar
outcomes can emerge through different pathways.

Our work extends Morse-Smale regression \cite{gerber2013morse} through robust
conditional expectation estimation on density-aware Riemannian graphs,
systematic treatment of spurious extrema through prominence filtering and basin
overlap clustering, path monotonicity validation to resolve long edge artifacts,
and co-monotonicity measures as alternatives to linear models within cells.
Beyond these technical advances, our introduction of association-based
decomposition through co-monotonicity complements Gerber's outcome-based
approach, providing a second lens for discovering structure.

The framework situates naturally within geometric data analysis rather than
topological data analysis. While TDA
\cite{carlsson2009topology,edelsbrunner2022computational} emphasizes topological
invariants like homology groups and persistence diagrams that remain unchanged
under continuous deformations, GDA focuses on geometric properties like
distances, angles, curvature, and geodesics that depend on the specific metric
structure. Our Riemannian graphs with their vertex and edge masses capture local
geometry that spectral methods can exploit for inference. The gradient flow we
compute respects this geometry through derivative weighting in co-monotonicity
coefficients and through the Riemannian metric implicit in the normalized
Laplacian. This geometric emphasis connects our work more closely to manifold
learning \cite{tenenbaum2000global,roweis2000nonlinear,coifman2006diffusion} and
spectral graph theory \cite{chung1997spectral,lovasz1993random} than to
persistent homology, though we acknowledge that the basin complex structure we
compute has topological aspects.

From a statistical perspective, our geometric decomposition provides data-driven
stratification for inference. Stratified sampling and stratified testing are
well-established ideas in statistics \cite{cochran1977sampling}, typically
requiring investigators to specify strata based on known covariates. Our
contribution is making stratification itself a geometric inference problem: the
strata emerge from graph structure through gradient flow or co-monotonicity
analysis rather than being predetermined. This connects to recent work on
conditional independence testing \cite{shah2020hardness} and context-specific
associations, but grounds these ideas in explicit geometric partitions of the
sample space.

The Bayesian inference framework we develop through posterior sampling on
weighted Laplacians provides credible intervals that respect the geometry of the
data manifold while enabling principled multiple testing corrections through the
geometric structure of co-monotonicity cells.

Section 2 establishes the mathematical framework of Riemannian graphs, functions
on graphs, and discrete gradient flow. Section 3 addresses the long edge problem
through path monotonicity validation and presents basin computation algorithms.
Section 4 introduces co-monotonicity coefficients, derives their properties, and
develops matrix extensions for multivariate analysis. Section 5 establishes
statistical inference procedures including vertex-wise permutation testing and
Bayesian credible intervals via posterior sampling. Section 6 presents the
geometric multiple testing framework through co-monotonicity cells and
biclustering. Section 7 discusses computational implementation in the gflow R
package. Section 8 concludes with discussion of limitations, relationships to
other methods, and future directions.

\section*{2. Mathematical Framework: Riemannian Graphs and Discrete Gradient Flow}

We begin with a question that motivates the entire geometric framework: given
observations $x_1, \ldots, x_n$ in $\mathbb{R}^d$ with associated outcomes $y_1,
\ldots, y_n$, how should we represent the inherent structure of the data in a way
that enables both robust estimation of conditional expectations and natural
stratification of the sample space? Standard approaches treat the data as points in
Euclidean space, but this perspective obscures the crucial geometric relationships
that determine how information should propagate between observations during inference.

\subsection*{2.1 From Point Clouds to Weighted Graphs}

Consider a simple example. Suppose we observe ten points arranged in two clusters in
the plane, with five points concentrated near $(0,0)$ and five near $(1,1)$.
Traditional kernel methods place a Gaussian at each point and sum, yielding a smooth
density estimate. Yet this approach treats all pairwise distances identically and
fails to recognize that points within clusters share local geometry while points in
different clusters do not, despite having similar distances to cluster boundaries.

The intersection k-nearest neighbor (ikNN) graph $G_{k}(X)$ associated with
observations $X \subset \mathbb{R}^{d}$ provides a more geometric perspective.
The vertices of $G_{k}(X)$ are all points $X$ and two vertices $x_{i}, x_{j}$
are connected by an edge if and only if
$\hat{N}_k(x_i) \cap \hat{N}_k(x_j) \neq \emptyset$, where $\hat{N}_k(x_i)$ is
the closed k-nearest neighbor ball of $x_i$, consisting of all k-nearest
neighbors of $x_i$ and the point $x_i$. For our two-cluster example with $k=4$,
edges connect points within each cluster but not across clusters (assuming
sufficient separation). The graph structure reveals the discrete geometry: the
two clusters appear as connected components, and any function defined on
vertices can diffuse within components but not between them.

However, unweighted graphs lose important information. In regions where points
lie densely, small Euclidean distances separate neighbors, while in sparse
regions, large distances occur. A single long edge might connect distant points
creating an artificial bridge that permits gradient flow to jump between
geometrically distant regions. We require vertex and edge weights that encode
both local density and geometric scale.

\subsection*{2.2 Riemannian Structure Through Mass Assignment}

We construct a Riemannian graph through systematic assignment of masses to
vertices and edges. The vertex mass $m_0(v)$ at vertex $v$ quantifies the local
concentration of observations, while edge masses $m_1(e)$ measure geometric
relationships between neighborhoods. These masses induce a complete metric
structure on the graph, enabling precise measurement of distances, angles, and
volumes.

Let $\mathcal{U} = \{\hat{N}_k(x_1), \ldots, \hat{N}_k(x_n)\}$ be the kNN balls
covering of $X = \{x_{1}, x_{2}, \ldots, x_{n}\}$. The nerve complex of this
covering has vertices corresponding to observations, with simplices recording
multi-way neighborhood intersections. We focus on the 1-skeleton (vertices and
edges), which forms a graph where an edge $[i,j]$ exists whenever
$\hat{N}_k(x_i) \cap \hat{N}_k(x_j) \neq \emptyset$. Thus, the intersection
k-nearest neighbor graph is the 1-skeleton of the nerve complex of the k-nearest
neighbor covering.

The vertex mass $m_0(i)$ represents the measure of the neighborhood $\hat{N}_k(x_i)$.
In practice, we use density-surrogate weights
\begin{equation}
m_0(i) = w(x_i) = (\varepsilon + d_k(x_i))^{-\alpha}
\end{equation}
where $d_k(x_i)$ denotes the distance from $x_i$ to its $k$-th nearest neighbor,
$\varepsilon > 0$ is a small regularization constant, and $\alpha \in [1,2]$
controls the degree of density weighting. The vertex masses are normalized so
that $\sum_{i=1}^n m_0(i) = n$.

The formula inverts neighborhood radius: points in dense regions have small
$d_k$ and receive large mass, while isolated points have large $d_k$ and receive
small mass. The exponent $\alpha$ modulates sensitivity to density variation.
Alternative formulations using exponential or rational kernels provide smooth
density dependence.

For an edge $e = [i,j]$, the edge mass $m_1(e)$ equals the total vertex mass in
the neighborhood intersection:
\begin{equation}
m_1([i,j]) = \sum_{\ell: x_\ell \in \hat{N}_k(x_i) \cap \hat{N}_k(x_j)} m_0(\ell).
\end{equation}

For two edges $e_{ij} = [i,j]$ and $e_{is} = [i,s]$ sharing vertex $i$, their
inner product equals the triple intersection mass:
\begin{equation}
\langle e_{ij}, e_{is} \rangle = \sum_{\ell: x_\ell \in \hat{N}_k(x_i) \cap \hat{N}_k(x_j) \cap \hat{N}_k(x_s)} m_0(\ell).
\end{equation}
Thus, $e_{ij}, e_{is}$ are orthogonal if and only if
$$
\hat{N}_k(x_i) \cap \hat{N}_k(x_j) \cap \hat{N}_k(x_s) = \emptyset
$$
assuming $m_{0}(x_{i}) > 0$ for all $x_{i}$.

These inner products encode geometric relationships beyond simple edge weights.
Two edges sharing a vertex are orthogonal if their corresponding neighborhoods
intersect trivially, while edges with large triple intersection mass form acute
angles. The collection of all edge inner products assembles into the edge mass
matrix $M_1$, with diagonal entries $M_1(e,e) = m_1(e)$ and off-diagonal entries
$M_1(e_{ij}, e_{is}) = \langle e_{ij}, e_{is} \rangle$.

We encode the vertex masses in a diagonal matrix
$M_0 = \text{diag}(m_0(1), \ldots, m_0(n))$. The pair $(M_0, M_1)$ constitutes
the Riemannian structure, determining inner products on the spaces of vertex
chains $C_0$ and edge chains $C_1$.

\subsection*{2.3 The Graph Laplacian and Diffusion}

The Riemannian structure determines a graph Laplacian that governs diffusion on
the weighted graph. We begin with the boundary operator $B_1: C_1 \to C_0$,
which maps edge chains to vertex chains according to
$$
(B_1 \alpha)(i) = \sum_{j \in N(i)} (\alpha([j,i]) - \alpha([i,j])),
$$
where $N(i)$ denotes the neighborhood of vertex $i$ and $[i,j]$ denotes the
oriented edge from $i$ to $j$. This operator encodes the graph's combinatorial
structure through the vertex-edge incidence relations.

We define the graph Laplacian using edge conductances derived from the edge mass
matrix. For each edge $e$, we set the conductance $c_e = m_1(e)$, using the edge
mass directly. Since the base measure
$m_0(i) = (\varepsilon + d_k(x_i))^{-\alpha}$ already inverts local scales,
edges in dense regions naturally receive large conductances while edges in
sparse regions receive small conductances. Assembling these into a diagonal
matrix $C = \text{diag}(c_1, \ldots, c_m)$, we construct the unnormalized
Laplacian
$$
L_{\text{div}} = B_1 C B_1^T.
$$
This operator is symmetric and positive semidefinite, with eigenvalues encoding
the graph's connectivity structure.

For computational purposes, we work with the symmetrized normalized Laplacian
$$
L_{\text{norm}} = M_0^{-1/2} L_{\text{div}} M_0^{-1/2},
$$
which has eigenvalues in $[0,2]$ and admits spectral decomposition through
standard symmetric eigensolvers. The normalization by vertex masses balances the
influence of vertices with different local densities, preventing high-degree
vertices from dominating the diffusion process.

The heat equation $\partial \rho / \partial t = -L_{\text{div}} \rho$ describes
how vertex masses evolve under diffusion. The solution
$\rho(t) = \exp(-tL_{\text{div}}) \rho(0)$ applies the heat kernel to the
initial distribution. For small $t$, diffusion smooths local irregularities
while preserving global structure; for large $t$, all mass flows toward the
stationary distribution. The Riemannian structure determines diffusion rates:
edges with large mass (small Riemannian length) facilitate rapid exchange, while
edges with small mass (large Riemannian length) impede flow.

The edge mass matrix $M_1$ contains not only diagonal entries $m_1(e)$ but also
off-diagonal entries encoding inner products between edges sharing vertices
through triple neighborhood intersections. In the Laplacian construction above,
we use only the diagonal structure for computational efficiency. The full
Riemannian geometry, including these off-diagonal terms, is incorporated in the
gradient computation used for basin analysis (Section 3), where it plays a
crucial role in determining flow directions. Extending the Laplacian
construction to utilize the complete $M_1$ structure remains an important
direction for future work.

\subsection*{2.4 Functions on Graphs and the Discrete Gradient}

We consider real-valued functions $f: V \to \mathbb{R}$ defined on graph
vertices. Such functions represent observed outcomes, fitted predictions, or any
other vertex-associated quantities. The gradient of $f$ measures how the
function changes across edges.

The gradient operator $\nabla f: E \to \mathbb{R}$ assigns to each oriented edge
a value encoding the directional rate of change of $f$. The Riemannian structure
determines this assignment through the adjoint of the boundary operator.
Formally, the gradient is defined by the relation
$$
\langle B_1 \phi, f \rangle_{M_0} = \langle \phi, \nabla f \rangle_{M_1}
$$
for all edge functions $\phi$ and vertex functions $f$, where the left side uses
the vertex inner product weighted by $M_0$ and the right side uses the edge
inner product weighted by $M_1$. This yields $\nabla f = B_1^*(f)$, where
$B_1^*: C_0 \to C_1$ denotes the adjoint operator.

In the diagonal case, where edges are orthogonal in the Riemannian metric
(meaning $M_1$ is diagonal with $M_1(e_{ij}, e_{is}) = 0$ for distinct edges
sharing a vertex), the gradient has an explicit formula. For an edge $e = [i,j]$
directed from vertex $i$ to vertex $j$, we have
$$
(\nabla f)(e) = \frac{m_0(j) f(j) - m_0(i) f(i)}{m_1(e)},
$$
where the vertex masses in the numerator weight the function values according to
local density, and the edge mass in the denominator normalizes by the geometric
scale of the neighborhood intersection.

This formula admits a natural interpretation. In regions where both vertices
have equal mass (uniform density), the gradient reduces to the function
difference divided by edge mass, analogous to a directional derivative where
edge mass acts as effective distance. In non-uniform regions, vertices with
larger mass contribute more strongly, reflecting that these vertices represent
denser neighborhoods where the function value carries greater statistical
weight. The division by edge mass ensures that the gradient measures rate of
change per unit geometric distance rather than absolute difference.

When the full non-diagonal structure of $M_1$ is employed (incorporating
off-diagonal entries from triple neighborhood intersections), computing
$\nabla f = B_1^*(f)$ requires solving the linear system $M_1 x = B_1^T M_0 f$,
which we address through iterative methods in the basin analysis of Section 3.

\subsection*{2.5 Spectral Filtering and Conditional Expectation Estimation}

Given observed outcomes $y_1, \ldots, y_n$, we estimate the conditional
expectation $E[Y|X]$ through spectral filtering on the graph Laplacian. The
empirical vertex function $y = (y_1, \ldots, y_n)$ contains both signal (the
true conditional expectation) and noise. Spectral methods decompose $y$ into
components corresponding to different geometric frequencies on the graph.

We begin with the eigendecomposition $L_{\text{norm}} = V \Lambda V^T$, where $\Lambda$
is diagonal with eigenvalues $0 = \lambda_1 \leq \lambda_2 \leq \cdots \leq \lambda_n \leq 2$
and $V$ contains the corresponding eigenvectors. The eigenvector $v_1$ associated with
$\lambda_1 = 0$ is constant (proportional to $M_0^{1/2} \mathbf{1}$), while eigenvectors
associated with small positive eigenvalues vary slowly across edges, and eigenvectors
with large eigenvalues oscillate rapidly.

The smoothed estimate takes the form
\begin{equation}
\hat{y} = \sum_{i=1}^n h(\lambda_i) \langle y, v_i \rangle v_i,
\end{equation}
where $h: [0,2] \to [0,1]$ is a spectral filter that attenuates high-frequency components.
We employ the heat kernel filter $h(\lambda) = \exp(-t\lambda)$, which corresponds to
solving the heat equation for time $t$:
\begin{equation}
\hat{y} = \exp(-tL_{\text{norm}}) y.
\end{equation}

The diffusion time $t$ controls smoothness: small $t$ yields estimates close to the
empirical values, while large $t$ produces heavily smoothed estimates. We select $t$
through generalized cross-validation or by monitoring convergence of the gradient flow
structure: we increase $t$ until the number and prominence of local extrema stabilize.

This spectral approach has several advantages over local smoothing methods. It naturally
adapts to the graph geometry, diffusing rapidly within well-connected regions and slowly
across sparse connections. It respects the Riemannian structure encoded in $M_0$ and
$M_1$, ensuring that smoothing follows the intrinsic manifold rather than the ambient
Euclidean space. It enables efficient computation through sparse matrix methods and
iterative eigensolvers, scaling to graphs with thousands of vertices.

\section*{3. The Long Edge Problem and Path Monotonicity Validation}

Real data-derived graphs pose a fundamental challenge for discrete gradient flow
computation: the k-nearest neighbor construction that makes the graph
computationally tractable simultaneously introduces edges that violate the
geometric principles underlying gradient trajectories. We confront the question
of why naive gradient flow fails on realistic graphs and develop a path-based
validation criterion that restores geometric faithfulness without requiring
expensive global computations.

\subsection*{3.1 The Challenge of Long Edges}

Consider constructing a k-nearest neighbor graph from a sample drawn from a
low-dimensional manifold embedded in high-dimensional space. When the manifold
curves or has boundaries, points that are close in Euclidean distance may lie
far apart along the manifold. A point near the edge of a curved region might
have as its k-th nearest neighbor a point on the opposite side of a valley,
leading to an edge that shortcuts across the manifold rather than following its
intrinsic geometry.

These long edges are necessary for graph connectivity. Without them, the graph
fragments into components corresponding to dense regions, preventing paths
between different parts of the sample. Yet these same edges can create artifacts
in gradient flow computation. Consider a vertex $v$ in a descending basin of a
local minimum $m$, where the smoothed function $\hat{y}$ decreases monotonically
along all paths through neighboring vertices toward $m$. If $v$ has a long edge
to a vertex $u$ in a different basin where $\hat{y}(u) < \hat{y}(v)$, the naive
gradient rule (follow the neighbor with minimal function value) directs the
trajectory to jump across basins, violating the fundamental property that
gradient flow should follow monotone paths along the underlying geometry.

\subsection*{3.2 Inadequacy of Existing Approaches}

One might attempt to resolve this issue through more sophisticated gradient
computation. For basin analysis, we can compute the discrete gradient
incorporating the full Riemannian structure via
$$
(\nabla y)_e = [M_1^{-1} B_1^T M_0 y]_e,
$$
where $B_1: \mathbb{R}^E \to \mathbb{R}^V$ is the boundary operator, $M_0$ is
the vertex mass matrix, and $M_1$ is the full edge mass matrix including
off-diagonal entries encoding geometric relationships between edges sharing
vertices (see Appendix). This gradient accounts for global geometric structure
beyond local function differences.

However, even this sophisticated gradient computation does not resolve the long
edge problem. The gradient formula incorporates edge masses that quantify
neighborhood overlaps, but the fundamental issue remains: a long edge
connecting vertices in different basins can still produce a gradient that
directs flow across basin boundaries, even when the edge has appropriately small
mass. The problem is not that the gradient is computed incorrectly, but that the
graph topology itself contains edges that violate the manifold geometry. No
local gradient computation can distinguish between a long edge that shortcuts
across a valley (problematic) versus one that legitimately connects distant
points along a genuine gradient path (acceptable).

Alternatively, one might restrict gradient flow to edges shorter than a quantile
threshold $\tau_q$, where $\tau_q$ is the $q$-quantile of the edge length
distribution for some $q \in [0.5, 1]$. This heuristic removes the most
problematic edges but discards potentially valid geometric information. A long
edge might legitimately connect vertices that lie on a genuine gradient
trajectory if the underlying function varies smoothly between them. Uniform
length thresholding cannot distinguish between geometrically faithful long edges
and artifactual ones, leading to unnecessary loss of connectivity and potential
fragmentation of basins.

What we need instead is a criterion that validates each edge based on the actual
function behavior along paths, not merely on edge length or local gradient
values. This motivates our path monotonicity approach.

\subsection*{3.3 Path Monotonicity as a Validation Criterion}

We return to the defining property of gradient trajectories in smooth Morse
theory. Let $f: M \to \mathbb{R}$ be a smooth function on a Riemannian manifold
$M$, and let $\gamma: [0,1] \to M$ be a gradient flow trajectory with
$\gamma(0) = p$ and $\gamma(1) = q$. For ascending flow,
$\gamma'(t) = +\nabla f(\gamma(t))$; for descending flow,
$\gamma'(t) = -\nabla f(\gamma(t))$. In both cases, the function varies
monotonically along $\gamma$. For ascending trajectories:
$$
\frac{d}{dt} f(\gamma(t)) = \langle \nabla f(\gamma(t)), \gamma'(t) \rangle = \|\nabla f(\gamma(t))\|^2 > 0,
$$
while for descending trajectories, the derivative has the opposite sign. The key
property is that $f$ changes monotonically along the entire path, assuming
$\gamma$ does not pass through critical points. This monotonicity is not merely
a local property at each point but a global constraint on the trajectory.

In the discrete setting, we translate this insight into an operational
criterion. For an edge $e = [i,j]$ to participate in gradient flow, we require
not only that the function increases from $i$ to $j$ (for ascending flow) but
that the increase reflects genuine geometric variation rather than artificial
jumping across the manifold. We validate this by examining alternative paths: if
the function truly varies smoothly in the region containing $i$ and $j$, then
the shortest path connecting them should exhibit consistent monotonic behavior.

Let $\gamma = (v_0, v_1, \ldots, v_k)$ be a path in $G$ with $v_0 = i$ and
$v_k = j$. The function $\hat{y}$ is monotone along $\gamma$ if for all
$\ell \in \{0, 1, \ldots, k-1\}$ we have $\hat{y}(v_{\ell+1}) > \hat{y}(v_\ell)$
for ascending monotonicity, or $\hat{y}(v_{\ell+1}) < \hat{y}(v_\ell)$ for
descending monotonicity. We quantify the degree of monotonicity through the
co-monotonicity coefficient along the path:
$$
\text{cm}(\hat{y}; \gamma) = \frac{\sum_{\ell=0}^{k-1} w_{v_\ell v_{\ell+1}} \Delta_{v_\ell v_{\ell+1}} \hat{y}}{\sum_{\ell=0}^{k-1} w_{v_\ell v_{\ell+1}} |\Delta_{v_\ell v_{\ell+1}} \hat{y}|},
$$
where
$\Delta_{v_\ell v_{\ell+1}} \hat{y} = \hat{y}(v_{\ell+1}) - \hat{y}(v_\ell)$ and
$w_{v_\ell v_{\ell+1}}$ are edge weights. $\text{cm}(\hat{y}; \gamma) = 1$ for
$\hat{y}$ strictly ascending along $\gamma$ and
$\text{cm}(\hat{y}; \gamma) = -1$ for $\hat{y}$ strictly descending along
$\gamma$.

The path co-monotonicity coefficient aggregates signed differences along the
path, normalized by total variation. Perfect monotonicity yields
$\text{cm}(\hat{y}; \gamma) = \pm 1$, while oscillating functions produce values
closer to zero. We employ this coefficient to define a validation procedure for
long edges.

For an edge $e = [i,j]$ with edge length $\Delta_e$ exceeding a threshold
$\tau_q$ (typically the $q$-quantile of edge lengths for $q \in [0.75, 0.90]$),
we compute the shortest path $\gamma_{[i,j]}$ from $i$ to $j$ in the graph
$G \setminus \{e\}$ (after temporarily removing edge $e$). If no such path
exists, the edge is essential for connectivity and we accept it by default. If
$\gamma_{[i,j]}$ exists, we evaluate $\text{cm}(\hat{y}; \gamma_{[i,j]})$ and
compare against a threshold $\theta \in [0.85, 1.0]$. For ascending flow, we
require $\text{cm}(\hat{y}; \gamma_{[i,j]}) \geq \theta$; for descending flow,
$\text{cm}(\hat{y}; \gamma_{[i,j]}) \leq -\theta$. The edge participates in
gradient flow only if this monotonicity condition holds.

Given a function $\hat{y}: V \to \mathbb{R}$, a quantile threshold $q$, and a monotonicity threshold $\theta$, the validated edge set for ascending flow is
$$
E_{\text{val}}^{\uparrow} = \{[i,j] \in E : \hat{y}(j) > \hat{y}(i) \text{ and } (\Delta_e \leq \tau_q \text{ or } \mathcal{V}^{\uparrow}(e))\},
$$
where $\mathcal{V}^{\uparrow}(e)$ holds if either no path from $i$ to $j$ exists in $G \setminus \{e\}$, or the shortest such path $\gamma_{[i,j]}$ satisfies $\text{cm}(\hat{y}; \gamma_{[i,j]}) \geq \theta$. Similarly, for descending flow:
$$
E_{\text{val}}^{\downarrow} = \{[i,j] \in E : \hat{y}(j) < \hat{y}(i) \text{ and } (\Delta_e \leq \tau_q \text{ or } \mathcal{V}^{\downarrow}(e))\},
$$
where $\mathcal{V}^{\downarrow}(e)$ holds if either no path from $i$ to $j$ exists in $G \setminus \{e\}$, or $\text{cm}(\hat{y}; \gamma_{[i,j]}) \leq -\theta$. We compute gradient trajectories using only edges in $E_{\text{val}}^{\uparrow}$ for ascending basins and $E_{\text{val}}^{\downarrow}$ for descending basins.

This criterion captures the essential geometric property of gradient
trajectories while remaining computationally tractable. Computing the shortest
path in $G \setminus \{e\}$ requires breadth-first search, an $O(|E|)$ operation
for sparse graphs. Evaluating path monotonicity involves a single pass through
the path vertices, adding negligible overhead. The total cost of validating all
long edges is $O(L \cdot |E|)$ where $L$ is the number of long edges, typically
a small fraction of $|E|$ for well-chosen $q$.

\subsection*{3.4 Local Extrema and Spurious Feature Removal}

We return to a fundamental question that arises whenever discrete Morse theory
is applied to finite samples from continuous manifolds: how should we
distinguish genuine features of the underlying geometry from artifacts
introduced by discretization and statistical estimation? The gradient flow
structure computed from path-validated edges naturally partitions the graph into
basins of attraction, with each basin containing vertices that flow to a common
local extremum. However, not all local extrema identified by this purely
combinatorial criterion represent meaningful features of the conditional
expectation function. Some extrema arise from sampling variability, others from
numerical artifacts in the smoothing process, and still others from the
discretization itself, where the restriction of a smooth function to graph
vertices inevitably creates spurious critical points that do not correspond to
extrema of the continuous function.

Consider first the geometric origin of spurious extrema. Let
$f: M \to \mathbb{R}$ be a smooth function on a Riemannian manifold $M$, and let
$X = \{x_1, \ldots, x_n\}$ be a finite sample from $M$. The restriction
$y = f|_X$ defines a function on the sample points, which we represent as a
function on the vertices of the graph $G_k(X)$ derived from the k-nearest
neighbor construction. Even when $f$ has only a small number of critical points
on $M$, the discrete function $y$ typically exhibits numerous local extrema on
$G_k(X)$. This phenomenon occurs because discrete extrema are defined purely by
local comparisons: a vertex $v$ is a local minimum if $y(v) < y(u)$ for all
neighbors $u \in N(v)$, and a local maximum if $y(v) > y(u)$ for all
$u \in N(v)$. The graph structure imposes a specific neighborhood system that
may not align with the natural neighborhoods in the continuous manifold, leading
to vertices that satisfy the discrete extremum condition despite lying on smooth
portions of the function landscape.

After spectral smoothing to estimate the conditional expectation, additional
spurious extrema can emerge from the interplay between the smoothing operator
and the graph geometry. The heat kernel filter $\exp(-tL_{\text{norm}})$
diffuses function values across edges, with diffusion rates determined by the
Riemannian structure. In regions where the graph locally approximates the
manifold geometry well, smoothing faithfully reconstructs the underlying smooth
function. However, in regions where long edges bridge across valleys or where
local density variations create anomalous mass distributions, the smoothed
function may develop local extrema that reflect these geometric artifacts rather
than true features of the conditional expectation.

We address this challenge through a systematic refinement pipeline that
progressively removes spurious extrema based on multiple geometric and
statistical criteria. The refinement process operates in stages, each targeting
a different source of spuriousness while preserving the structure of the
function landscape.

The first filtering stage removes extrema whose function values are
insufficiently distinct from the global mean. We compute the relative value
$r(v) = \hat{y}(v) / \bar{y}$ for each extremum vertex $v$, where
$\bar{y} = n^{-1} \sum_{i=1}^n y(i)$ denotes the mean of $y$ across all
vertices. For local maxima, we retain only those satisfying
$r(v) \geq \rho_{\max}$ for a threshold $\rho_{\max} > 1$ (typically
$\rho_{\max} \in [1.1, 1.5]$), while for local minima we require
$r(v) \leq \rho_{\min}$ with $\rho_{\min} < 1$ (typically
$\rho_{\min} \in [0.5, 0.9]$). These thresholds focus subsequent analysis on
prominent features that rise substantially above or descend significantly below
the average function level. Extrema failing these criteria represent minor
fluctuations that, while formally satisfying the local comparison condition,
lack sufficient magnitude to warrant interpretation as distinct features of the
conditional expectation surface.

The second stage addresses redundancy arising from multiple nearby extrema
representing the same underlying feature. We quantify similarity between extrema
through their basin overlap. For two basins $A_i$ and $A_j$ with vertex sets
$V_i$ and $V_j$, the overlap coefficient is defined as
\begin{equation}
\omega(A_i, A_j) = \frac{|V_i \cap V_j|}{\min(|V_i|, |V_j|)},
\end{equation}
which measures what fraction of the smaller basin's vertices are shared with the
larger basin. This asymmetric measure emphasizes cases where one basin is
substantially contained in another, indicating that the corresponding extrema
likely represent the same feature at different scales of resolution. We
construct an overlap graph where vertices correspond to extrema of the same type
(all maxima or all minima), and we add an edge between extrema $i$ and $j$
whenever $\omega(A_i, A_j) \geq \omega_{\text{thld}}$ for a threshold
$\omega_{\text{thld}} \in [0.10, 0.20]$. Connected components in this overlap
graph identify clusters of similar extrema. Within each cluster, we merge the
basins by retaining only the extremum with the most extreme function value
(highest for maxima, lowest for minima) and assigning all vertices in the
cluster's combined basin to this representative extremum. This consolidation
reduces redundancy while preserving the essential gradient flow structure, as
vertices in merged basins still flow toward genuine extrema, just through a
simplified representative structure.

The third stage removes extrema whose basins exhibit geometric characteristics
suggesting isolation artifacts or boundary effects. We compute two complementary
measures of basin geometry for each extremum. The first is the mean hop-k
distance, which quantifies extended neighborhood isolation. For an extremum at
vertex $v$, we identify all vertices $U_k(v)$ at graph distance exactly $k$ from
$v$ through breadth-first search, then compute the mean geodesic distance
\begin{equation}
d_k(v) = \frac{1}{|U_k(v)|} \sum_{u \in U_k(v)} d_G(v,u),
\end{equation}
where $d_G(v,u)$ denotes the length of the shortest path from $v$ to $u$ in the
metric graph. This measure captures whether the extremum lies in a locally
sparse or dense region at scale $k$ (typically $k=2$ or $k=3$). High values of
$d_k(v)$ indicate that vertices at hop distance $k$ are geometrically far from
$v$, suggesting the extremum sits in an isolated or boundary region.

The second geometric measure is the effective degree $\deg_{\text{eff}}(v)$,
defined as the sum of density-surrogate weights over all neighbors:
\begin{equation}
\deg_{\text{eff}}(v) = \sum_{u \in N(v)} w(e_{vu}),
\end{equation}
where $w(e_{vu})$ denotes the edge weight (often taken as the edge mass
$m_1([v,u])$). This weighted degree quantifies how well-connected the vertex is
to the graph structure, accounting for local density variations. Vertices with
anomalously low effective degree relative to the global distribution may
represent poorly sampled boundary regions where the discrete gradient flow is
unreliable.

We convert both geometric measures to percentile ranks across all vertices: for
$d_k$, we compute
$$
p_k(v) = |\{u : d_k(u) \leq d_k(v)\}| / n,
$$
which gives the fraction of vertices with hop-k distance at most that of $v$.
Similarly, for effective degree, we compute
$$
p_{\deg}(v) = |\{u : \deg_{\text{eff}}(u) \geq \deg_{\text{eff}}(v)\}| / n,
$$
giving the fraction of vertices with effective degree at least that of $v$ (note
the reversed inequality, as high degree indicates good connectivity). We retain
extrema only if both geometric measures fall within acceptable ranges: typically
$p_k(v) < 0.90$ and $p_{\deg}(v) > 0.10$, though these thresholds may be
adjusted based on the specific graph geometry and application requirements.

The complete refinement pipeline applies these stages sequentially: relative
value filtering removes extrema with insufficient magnitude, overlap-based
clustering and merging consolidates redundant features, and geometric filtering
removes isolated or poorly-connected extrema. Each stage preserves the basin
structure for retained extrema, so the final refined basin structure maintains
complete information about gradient flow for all vertices assigned to surviving
extrema. This multi-stage approach balances statistical and geometric
considerations, ensuring that the final set of extrema represents genuine
features of the estimated conditional expectation function rather than artifacts
of discretization, sampling, or smoothing.

The filtering criteria involve several tunable parameters: the relative value
thresholds $\rho_{\max}$ and $\rho_{\min}$, the overlap threshold
$\omega_{\text{thld}}$, and the geometric percentile thresholds for $p_k$ and
$p_{\deg}$. In practice, these parameters can be selected through exploratory
analysis, examining how the number and prominence of retained extrema vary with
parameter choices, or through cross-validation by assessing the stability of the
resulting basin structure under perturbations of the smoothed function.
Conservative choices (stricter thresholds) produce fewer extrema with higher
confidence, while liberal choices preserve more potential features at the cost
of retaining some spurious extrema. The modular design of the pipeline allows
each filtering stage to be enabled or disabled independently, providing
flexibility for different application contexts and data characteristics.

\subsection*{3.5 Basin Computation with Validated Gradient Flow}

We turn now to the computational construction of basins of attraction using
validated gradient flow. The basin structure provides the foundation for
regional statistical inference, partitioning the sample space into regions where
outcomes flow toward common extrema. Before developing the discrete algorithm,
we recall the classical construction from smooth Morse theory, which both
motivates our approach and highlights a fundamental difference between the
continuous and discrete settings that has important implications for statistical
applications.

\subsubsection*{Classical Basins and Gradient Flow Trajectories}

In the smooth setting, consider a Morse function $f: M \to \mathbb{R}$ on a
compact Riemannian manifold $M$ without boundary. The gradient flow generates
two types of trajectories from any regular point $x \in M$ (a point where
$\nabla f(x) \neq 0$). The ascending trajectory satisfies the differential
equation $\gamma'(t) = +\nabla f(\gamma(t))$ with initial condition
$\gamma(0) = x$, following the direction of steepest increase until reaching a
local maximum as $t \to \infty$. The descending trajectory satisfies
$\gamma'(t) = -\nabla f(\gamma(t))$, flowing downhill to a local minimum. The
Morse-Smale condition (generic transversality of stable and unstable manifolds)
ensures that these trajectories are well-defined and unique for almost all
starting points, with trajectories terminating at critical points rather than
wandering indefinitely.

These trajectories induce maps $\pi_{\uparrow}: M \to \text{Max}(f)$ and
$\pi_{\downarrow}: M \to \text{Min}(f)$ that assign to each point the terminus
of its ascending and descending flows. For a local minimum $m$, the ascending
basin (or unstable manifold) is defined as
\begin{equation}
A(m) = \{x \in M : \pi_{\downarrow}(x) = m\},
\end{equation}
containing all points whose descending flow terminates at $m$. For a local
maximum $M$, the descending basin (or stable manifold) is
\begin{equation}
D(M) = \{x \in M : \pi_{\uparrow}(x) = M\},
\end{equation}
containing all points whose ascending flow terminates at $M$. The terminology
reflects the direction of flow reaching the critical point: vertices in $A(m)$
descend to reach $m$, while vertices in $D(M)$ ascend to reach $M$.

A crucial property of the smooth setting is that basins of the same type are
disjoint. No point can have descending flow terminating at two distinct minima,
nor can ascending flow from a single point reach two distinct maxima. This
uniqueness follows from the smooth dependence of gradient trajectories on
initial conditions and the Morse-Smale transversality assumption. The ascending
basins $\{A(m) : m \in \text{Min}(f)\}$ partition the manifold into disjoint
regions, as do the descending basins $\{D(M) : M \in \text{Max}(f)\}$. The
intersections $C(m,M) = A(m) \cap D(M)$ define gradient flow cells, which tile
the manifold into regions where both ascending and descending flows have unique
termini.

\subsubsection*{Non-Uniqueness in the Discrete Setting}

When we discretize the gradient flow by restricting to a finite graph
$G = (V,E)$, this uniqueness property fails. The discrete gradient flow is
defined by following edges of steepest ascent or descent, but when multiple
neighbors have identical or nearly identical function values, the choice of
which edge to follow becomes ambiguous. More fundamentally, the graph topology
itself creates situations where multiple distinct paths exist between a vertex
and an extremum, each exhibiting monotonic function behavior, yet leading to
different flow trajectories depending on the algorithmic tie-breaking rules.

Consider a simple example that illustrates this phenomenon. Let $G$ be a star
graph with a central vertex $c$ and three arms, each consisting of a single edge
connecting $c$ to a terminal vertex. Label the terminal vertices $v_1$, $v_2$,
and $v_3$. Define a function $\hat{y}$ with values $\hat{y}(v_1) = 0$,
$\hat{y}(v_2) = 0$, $\hat{y}(c) = 1$, and $\hat{y}(v_3) = 2$. The vertices $v_1$
and $v_2$ are both local minima (each has function value less than its only
neighbor $c$), while $v_3$ is the unique local maximum.

Now consider the ascending basins. The vertex $c$ has function value
intermediate between the two minima and the maximum. From $c$, descending flow
could reasonably proceed to either $v_1$ or $v_2$, as both are neighbors with
lower function values. If we break ties by choosing the neighbor with minimal
function value, both $v_1$ and $v_2$ qualify equally. Different tie-breaking
rules lead to different basin assignments: one rule might assign $c$ to
$A(v_1)$, another to $A(v_2)$. Regardless of the choice, the resulting ascending
basins are not disjoint. If $c \in A(v_1)$, then since $v_2$ is a minimum, we
have $A(v_2) = \{v_2\}$. But if we run the basin construction algorithm
symmetrically from both minima using breadth-first exploration (as we describe
below), the vertex $c$ appears reachable from both $v_1$ and $v_2$ through
monotone descending paths ($v_1 \to c$ and $v_2 \to c$ both ascend, so tracing
backward from $c$ can descend to either minimum). Thus $c$ potentially belongs
to both ascending basins, creating an intersection $A(v_1) \cap A(v_2) \ni c$.

This non-uniqueness has important statistical implications. In the continuous
setting, the partition into gradient flow cells provides an unambiguous spatial
decomposition for regional inference. In the discrete setting, ambiguous
vertices near basin boundaries require explicit resolution through tie-breaking
rules or probabilistic assignment. Our algorithm addresses this by using
breadth-first exploration to establish basin membership definitively based on
discovery order, but we acknowledge that alternative algorithmic choices could
produce different basin structures for ambiguous vertices. This sensitivity to
algorithmic details suggests that statistical inference should focus on vertices
deep within basins, where flow direction is unambiguous, rather than on vertices
near boundaries where multiple extrema exert comparable influence.

\subsubsection*{Backward Tracing Through Breadth-First Exploration}

We construct basins through breadth-first exploration that works backward along
flow trajectories. The key insight is that to find all vertices whose ascending
flow reaches a maximum $M$, we start at $M$ itself and explore outward,
accepting vertices that lie "below" already-accepted vertices in function value.
This backward tracing ensures that every accepted vertex has a monotone
increasing path to $M$, which corresponds precisely to the forward ascending
trajectory from that vertex terminating at $M$.

Consider the descending basin $D(M) = \{v : \pi_{\uparrow}(v) = M\}$ of a local
maximum $M$. We initialize the basin with $D(M) = \{M\}$ and a queue containing
$M$. We maintain a distance map $\text{dist}[v]$ recording the hop distance from
$M$ to each accepted vertex, and a predecessor map $\text{pred}[v]$ tracking the
path back to $M$. The algorithm proceeds iteratively: while the queue is
non-empty, we dequeue a vertex $u \in D(M)$ and examine each neighbor
$v \notin D(M)$. If $\hat{y}(v) < \hat{y}(u)$, then the edge $[v,u]$ represents
a valid step in an ascending trajectory from $v$ toward $M$ (since moving from
$v$ to $u$ increases the function value). To verify that this edge participates
in genuine gradient flow rather than an artifactual jump across basins, we apply
the validation criterion from Section 3.3.

For edges with length $\Delta_{uv} \leq \tau_q$ (below the quantile threshold),
we accept $v$ immediately, as short edges are presumed to follow local manifold
geometry faithfully. For long edges with $\Delta_{uv} > \tau_q$, we compute the
shortest alternative path $\gamma_{[u,v]}$ from $u$ to $v$ in the graph
$G \setminus \{[u,v]\}$ (temporarily removing the direct edge). If no such path
exists, the edge is essential for connectivity and we accept it by default. If
an alternative path exists, we evaluate its co-monotonicity coefficient
$\text{cm}(\hat{y}; \gamma_{[u,v]})$. For descending exploration (where we seek
vertices with $\hat{y}(v) < \hat{y}(u)$ that can ascend to $M$), the path from
$u$ to $v$ should exhibit descending monotonicity (function values decreasing
along the path from $u$ toward $v$). We require
$\text{cm}(\hat{y}; \gamma_{[u,v]}) \leq -\theta$ for a threshold
$\theta \in [0.85, 1.0]$. If this condition holds, the long edge is validated
and we accept $v$ into the basin, setting $\text{dist}[v] = \text{dist}[u] + 1$,
$\text{pred}[v] = u$, and adding $v$ to the queue for further exploration. If
validation fails, we reject this particular edge from $u$ to $v$ and continue
examining other neighbors.

This backward exploration guarantees that every vertex in $D(M)$ is connected to
$M$ through a validated path where function values increase monotonically.
Vertices reachable from $M$ only through invalidated long edges remain outside
the basin, preventing spurious basin jumping while maintaining computational
efficiency. The validation occurs lazily during breadth-first search, so we
evaluate co-monotonicity only for long edges that actually arise in potential
basin membership, avoiding exhaustive validation of the entire graph.

\begin{algorithm}
\caption{Validated Descending Basin Computation}
\begin{algorithmic}
\STATE \textbf{Input:} Graph $G = (V, E)$, function $\hat{y}: V \to \mathbb{R}$, local maximum $M$, quantile threshold $q$, monotonicity threshold $\theta$
\STATE \textbf{Output:} Descending basin $D(M) = \{v : \pi_{\uparrow}(v) = M\} \subseteq V$
\STATE
\STATE Compute edge length quantile $\tau_q \leftarrow \text{quantile}(\{\Delta_e : e \in E\}, q)$
\STATE Initialize $D(M) \leftarrow \{M\}$, $\text{queue} \leftarrow [M]$, $\text{dist}[M] \leftarrow 0$, $\text{pred}[M] \leftarrow \text{null}$
\STATE
\WHILE{queue is not empty}
    \STATE $u \leftarrow \text{dequeue()}$
    \FOR{each neighbor $v$ of $u$ in $G$}
        \IF{$v \notin D(M)$ \textbf{and} $\hat{y}(v) < \hat{y}(u)$}
            \STATE $e \leftarrow [u,v]$
            \IF{$\Delta_e \leq \tau_q$}
                \STATE $\text{valid} \leftarrow \text{true}$
                \COMMENT{Short edge: $v$ can ascend through $u$ to $M$}
            \ELSE
                \STATE $\gamma_{[u,v]} \leftarrow \text{ShortestPath}(u, v, G \setminus \{e\})$
                \IF{$\gamma_{[u,v]} = \emptyset$}
                    \STATE $\text{valid} \leftarrow \text{true}$
                    \COMMENT{Edge essential for connectivity}
                \ELSE
                    \STATE $\text{valid} \leftarrow (\text{cm}(\hat{y}; \gamma_{[u,v]}) \leq -\theta)$
                    \COMMENT{Validate descending path monotonicity}
                \ENDIF
            \ENDIF
            \IF{valid}
                \STATE $D(M) \leftarrow D(M) \cup \{v\}$
                \STATE $\text{dist}[v] \leftarrow \text{dist}[u] + 1$
                \STATE $\text{pred}[v] \leftarrow u$
                \STATE enqueue($v$)
            \ENDIF
        \ENDIF
    \ENDFOR
\ENDWHILE
\STATE \textbf{return} $D(M)$, dist, pred
\end{algorithmic}
\end{algorithm}

Ascending basins for local minima follow the same algorithmic structure with
reversed monotonicity conditions. To construct
$A(m) = \{v : \pi_{\downarrow}(v) = m\}$ for a local minimum $m$, we initialize
with $A(m) = \{m\}$ and perform breadth-first exploration accepting neighbors
$v$ of vertices $u \in A(m)$ when $\hat{y}(v) > \hat{y}(u)$. This explores
upward from $m$ (in the ascending direction) while tracing descending flow
trajectories backward, identifying all vertices from which descending flow
reaches $m$. For long edges, we require
$\text{cm}(\hat{y}; \gamma_{[u,v]}) \geq \theta$, validating that the
alternative path from $u$ to $v$ exhibits ascending monotonicity consistent with
the upward exploration direction.

The complete partition of vertices into basins emerges from running this
algorithm for all detected local extrema. Due to the non-uniqueness discussed
above, some vertices may be discovered during exploration from multiple extrema
of the same type. We resolve such conflicts through tie-breaking rules based on
discovery order: the first extremum to reach a vertex during breadth-first
exploration claims that vertex for its basin. This deterministic rule ensures
each vertex belongs to exactly one ascending basin and exactly one descending
basin, though the specific assignment depends on the order in which extrema are
processed. Alternative tie-breaking strategies include assigning ambiguous
vertices based on minimal path length to extrema, maximal function value
difference along paths, or probabilistic assignment proportional to path
validation scores. The sensitivity of boundary vertex assignments to these
algorithmic choices reinforces that robust inference should focus on the
interior of basins rather than their boundaries.

For vertices that remain unassigned after processing all extrema, we have two
options. The first is to leave them as isolated singletons, which may occur for
vertices in flat regions where no clear gradient direction exists. The second is
to perform a final assignment phase where unassigned vertices are allocated to
the nearest basin based on graph distance, function value similarity, or other
proximity measures. In practice, validated gradient flow with appropriate
quantile and monotonicity thresholds produces nearly complete basin coverage,
with unassigned vertices arising primarily in highly ambiguous flat regions or
poorly connected boundary zones.

\subsubsection*{Computational Complexity and Practical Considerations}

The algorithm's computational cost decomposes into several components. The
initial quantile computation requires sorting all edge lengths, an
$O(|E| \log |E|)$ operation performed once before basin construction begins. For
each extremum, the breadth-first exploration visits each vertex at most once and
examines each incident edge at most once, giving $O(|V| + |E|)$ per extremum in
the absence of validation. When validation is required for a long edge
$e = [u,v]$, computing the shortest alternative path using breadth-first search
costs $O(|E|)$ in sparse graphs. Evaluating the co-monotonicity coefficient
along a path of length $\ell$ costs $O(\ell)$, typically $O(\log |V|)$ for paths
in graphs with good expansion properties.

The total validation cost depends on the quantile threshold $q$ and the graph
geometry. Choosing $q \in [0.75, 0.90]$ ensures that only the longest
$10$-$25$\% of edges require validation. In well-structured graphs where long
edges are rare and localized, validation overhead remains modest. In
pathological cases where many long edges connect distant regions, validation
costs can approach $O(L \cdot |E|)$ where $L$ is the number of long edges
examined during all basin explorations. For typical applications with
$n = 500$-$5000$ vertices and $k = 10$-$50$ nearest neighbors, basin computation
completes in seconds to minutes on modern hardware, with validation adding at
most a factor of $2$-$5$ overhead compared to naive gradient flow.

Memory requirements are dominated by storing the graph structure ($O(|E|)$ for
adjacency lists and edge lengths), the distance and predecessor maps for each
basin ($O(|V|)$ per extremum), and temporary storage for shortest path
computations ($O(|V|)$). For graphs with $n = 10^4$ vertices and moderate
degree, total memory consumption remains well under 1 GB, enabling in-memory
computation on standard workstations. Parallel computation across extrema is
straightforward, as basin constructions are independent until the final
tie-breaking phase, offering near-linear speedup on multi-core systems.

\section*{4. Co-Monotonicity Coefficients}

Having established gradient flow partitions through local extrema and their
basins, we turn to quantifying associations between functions on graph vertices.
The Pearson correlation coefficient provides the foundational framework for
measuring whether two variables vary together, serving as one of the first and
most widely used measures in theoretical statistics
\citep{pearson1895notes,pearson1896regression}. However, Pearson's formulation
treats all observations as equally related, computing association through
deviations from global means without reference to any spatial or relational
structure among observations. When data possess geometric organization encoded
by graphs or simplicial complexes, this structure-agnostic approach discards
valuable information about how associations vary across different regions of the
sample space. The fundamental question driving our development is thus: how can
we adapt correlation's proven normalization structure to respect geometric
relationships, yielding measures that detect directional concordance locally
while adapting to spatial heterogeneity in association patterns? We develop
co-monotonicity coefficients as context-dependent versions of classical
correlation, replacing global mean deviations with edge-based directional
concordance computed within local graph neighborhoods. This geometric refinement
preserves the interpretability and bounded range of correlation coefficients
while enabling vertex-level resolution of association structure.

\subsection*{4.1 Global Association Measures}

Consider an outcome function $y: V \to \mathbb{R}$ and a potential predictor
$z: V \to \mathbb{R}$ defined on the vertices of a graph $G = (V, E)$ derived from
high-dimensional data. Standard correlation analysis computes the Pearson coefficient
\begin{equation}
\rho(y, z) = \frac{\text{Cov}(y, z)}{\sigma_y \sigma_z}
= \frac{\sum_{v \in V} (y(v) - \bar{y})(z(v) - \bar{z})}
{\sqrt{\sum_{v \in V} (y(v) - \bar{y})^2} \sqrt{\sum_{v \in V} (z(v) - \bar{z})^2}},
\end{equation}
where $\bar{y}$ and $\bar{z}$ denote sample means. This measure treats all
vertex pairs equally, implicitly assuming that observations at vertices $v$ and
$u$ are as related as observations at $v$ and $w$, regardless of whether edges
connect these pairs in the graph. When the graph encodes meaningful proximity
relationships, as in spatial data or network analysis, this assumption discards
valuable structural information. In particular, when a feature $z$ varies
concordantly with $y$ in one region of the graph but discordantly in
another, these opposing local relationships may cancel in the global
correlation, yielding a coefficient near zero despite strong region-specific
associations.

Moreover, correlation measures linear association in terms of deviations from
means, which may not align with the geometric properties of interest. Within a
gradient flow cell where the outcome $y$ varies monotonically, we seek features
that exhibit monotone co-variation along the same paths, increasing when $y$
increases and decreasing when $y$ decreases. Mutual information addresses
nonlinearity and can be computed on graph structures, providing a viable
alternative for quantifying associations between vertex functions. However,
mutual information and other global measures produce a single summary of the
relationship between $y$ and $z$ across all vertices. They cannot reveal spatial
heterogeneity in association patterns, nor can they identify regions where the
relationship is strong versus weak, or positive versus negative. When
association structure varies across the graph, as commonly occurs in gradient
flow cells with different monotonic behaviors, a single global measure obscures
this regional variation. While mutual information can in principle be localized
to vertices or paths through kernel-weighted density estimation (see Appendix
B), such approaches introduce substantial practical challenges in estimation and
interpretation.

Co-monotonicity coefficients address this limitation by providing vertex-level
measures rather than global summaries. For each vertex $v$, we compute a
coefficient $c(y, z; v)$ that quantifies directional concordance between $y$ and
$z$ within the local neighborhood of $v$. This yields a function
$c: V \to [-1, 1]$ that maps vertices to association strengths, enabling
identification of regions where features exhibit strong positive co-variation,
regions with negative relationships, and regions where associations are weak or
absent. The coefficients have intuitive geometric interpretation: they measure
whether $y$ and $z$ tend to increase or decrease together across edges incident
to each vertex, directly capturing the directional concordance that
characterizes monotonic co-variation within gradient flow structures. This
vertex-level resolution proves essential for regional inference, where
statistical power comes from identifying coherent association patterns within
cells while allowing relationships to differ across cells.

We formalize these concepts through two complementary perspectives on
directional concordance. At each vertex $v \in V$, we define a coefficient
$c(y, z; v) \in [-1, 1]$ that measures whether $y$ and $z$ tend to increase or
decrease together across edges incident to $v$. This local measure naturally
extends to paths $\gamma = (\gamma_0, \gamma_1, \ldots, \gamma_n)$, where we
compute $c(y, z; \gamma)$ to quantify directional concordance along the sequence
of edges comprising $\gamma$. The vertex-based coefficients enable spatial
mapping of association patterns, while path-based coefficients characterize
monotonic behavior along gradient flow trajectories. We develop both
perspectives in the following sections, beginning with the vertex-centered
formulation.

\subsection*{4.2 Vertex-Level Co-Monotonicity Coefficients}

We seek vertex-level measures that quantify whether two functions increase or
decrease together across edges in local neighborhoods. The approach builds on
edge-wise directional concordance, aggregating signed products of function
changes across incident edges with appropriate normalization to yield
coefficients bounded in $[-1,1]$.

We begin by formalizing the notion of directional agreement along a single edge.
Let $e = [v,u] \in E$ be an edge connecting vertices $v$ and $u$, and define the
edge difference operator acting on a function $f: V \to \mathbb{R}$ by
\begin{equation}
\Delta_e f = f(u) - f(v).
\end{equation}
This operator measures the change in $f$ when moving from $v$ to $u$ along edge
$e$. The choice of orientation (which vertex to designate as initial point) is
arbitrary; what matters is that we consistently use the same orientation for all
functions when computing products of differences, ensuring that the product
captures directional concordance rather than being affected by orientation
choices.

The product $\Delta_e y \cdot \Delta_e z$ captures the directional concordance
of $y$ and $z$ along edge $e$. When both functions increase
($\Delta_e y > 0$ and $\Delta_e z > 0$) or both decrease
($\Delta_e y < 0$ and $\Delta_e z < 0$), the product is positive, indicating
co-monotonic behavior. When the functions change in opposite directions, the
product is negative, indicating anti-monotonic behavior. When at least one
function remains constant across the edge, the product vanishes.

To construct a vertex-level measure, we aggregate these products over edges
incident to each vertex. Let $N(v)$ denote the set of neighbors of $v$ (vertices
connected to $v$ by edges). A natural approach sums the weighted products
$w_e \Delta_e y \cdot \Delta_e z$ over all edges $e = [v,u]$ with $u \in N(v)$,
where $w_e \geq 0$ are edge weights. However, this raw sum depends on the scales
of $y$ and $z$, making comparisons across different function pairs problematic.
We address this by normalizing in a manner analogous to Pearson correlation,
dividing by the geometric mean of the total weighted squared changes in each
function separately. This normalization ensures the resulting coefficient lies in
$[-1,1]$ regardless of function scales, with extremal values indicating perfect
proportional co-variation or anti-variation.

Given functions $y, z: V \to \mathbb{R}$ on a weighted graph $G = (V, E)$ with
edge weights $w_e \geq 0$, the correlation-type co-monotonicity coefficient at
vertex $v$ is
\begin{equation}
\text{cm}_{\text{cor}}(y,z;w)(v) = \frac{\sum_{u \in N(v)} w_e \Delta_e y \cdot \Delta_e z}{\sqrt{\sum_{u \in N(v)} w_e (\Delta_e y)^2} \sqrt{\sum_{u \in N(v)} w_e (\Delta_e z)^2}},
\end{equation}
where $e = [v,u]$ denotes the edge connecting $v$ and $u$, and
$\Delta_e y = y(u) - y(v)$, $\Delta_e z = z(u) - z(v)$ are the edge differences
oriented away from $v$. When either denominator term vanishes (indicating that
one of the functions is constant across all edges incident to $v$), we define
$\text{cm}_{\text{cor}}(y,z;w)(v) = 0$ by convention. This formulation preserves
the correlation-type normalization structure of Pearson's coefficient while
replacing global mean deviations with local edge differences, yielding a
geometric refinement that respects graph structure.

The normalization by geometric mean of squared changes ensures that
$\text{cm}_{\text{cor}}(y,z;w)(v) \in [-1,1]$ for all vertices, with extremal
values achieved under perfect proportionality. If $\Delta_e z = k \Delta_e y$
for some constant $k > 0$ across all edges incident to $v$, then
$\text{cm}_{\text{cor}}(y,z;w)(v) = 1$. If $\Delta_e z = -k \Delta_e y$ for
$k > 0$, then $\text{cm}_{\text{cor}}(y,z;w)(v) = -1$. The coefficient vanishes
when positive and negative products balance in a squared-magnitude-weighted
sense.

A particularly important special case arises when edge weights are chosen as
$w_e = 1/\ell_e$ where $\ell_e$ denotes the length of edge $e$. This
derivative-weighted co-monotonicity coefficient takes the form
\begin{equation}
\text{cm}_{\text{cor},\partial}(y,z)(v) = \frac{\sum_{u \in N(v)} \frac{\Delta_e y}{\ell_e} \cdot \frac{\Delta_e z}{\ell_e}}{\sqrt{\sum_{u \in N(v)} \left(\frac{\Delta_e y}{\ell_e}\right)^2} \sqrt{\sum_{u \in N(v)} \left(\frac{\Delta_e z}{\ell_e}\right)^2}},
\end{equation}
where the ratios $\Delta_e y / \ell_e$ and $\Delta_e z / \ell_e$ approximate
directional derivatives of $y$ and $z$ along the edges. This form admits a
compelling geometric interpretation: on a smooth Riemannian manifold, the
coefficient measures the cosine of the angle between gradient vector fields,
providing theoretical justification for this normalization choice. We develop
this connection formally in Section 4.4, where we show that
$\text{cm}_{\text{cor},\partial}$ arises naturally from the Riemannian metric
structure on smooth manifolds.

Throughout the remainder of this paper, we adopt the simplified notation
$\text{cm}(y,z;w)(v)$ to denote the correlation-type co-monotonicity coefficient
$\text{cm}_{\text{cor}}(y,z;w)(v)$ when the normalization scheme is clear from
context. For derivative weighting, we write $\text{cm}_\partial(y,z)(v)$ as
shorthand for $\text{cm}_{\text{cor},\partial}(y,z)(v)$. The subscript "cor" is
retained only when explicitly contrasting with alternative normalizations.

\subsection*{4.3 Geometric Interpretation via Gradient Alignment}

We now establish a rigorous connection between correlation-type co-monotonicity
and angular correlation of gradients on smooth Riemannian manifolds. This
provides both theoretical justification for the derivative-weighted formulation
and intuitive geometric meaning for the coefficient values.

Suppose data $X = \{x_1, \ldots, x_n\}$ are sampled from a smooth
$d$-dimensional Riemannian manifold $M$ embedded in $\mathbb{R}^D$ for
$D \geq d$. Let $y, z: M \to \mathbb{R}$ be smooth functions with non-vanishing
gradients in a neighborhood of a point $x_v \in M$. The k-nearest neighbor graph
$G_n$ constructed from $X$ approximates $M$ in the limit of dense sampling, with
appropriate scaling $k = k(n)$ such that $k/n \to 0$ and $k/\log n \to \infty$
as $n \to \infty$.

For a vertex $v$ corresponding to point $x_v \in M$, the neighborhood $N(v)$
consists of the $k$ nearest sample points, which are approximately uniformly
distributed in a geodesic ball $B_{r_k}(x_v)$ of radius $r_k$ around $x_v$,
where $r_k$ is the distance to the $k$-th nearest neighbor. As $n \to \infty$,
we have $r_k \to 0$ with rate determined by the local density of the sampling
distribution and the scaling of $k$.

Consider an edge $e = [v,u]$ connecting $v$ to a neighbor $u$ at point
$x_u \in M$ with $d_M(x_v, x_u) = \Delta_e$, where $d_M$ denotes the Riemannian
distance on $M$. For small $\Delta_e$, the function difference can be
approximated by the directional derivative:
\begin{equation}
\Delta_e y = y(x_u) - y(x_v) = \langle \nabla_M y(x_v), \xi_e \rangle \Delta_e + O(\Delta_e^2),
\end{equation}
where $\xi_e \in T_{x_v}M$ is the unit tangent vector at $x_v$ pointing in the
direction of the geodesic from $x_v$ to $x_u$, and $\nabla_M y$ denotes the
Riemannian gradient of $y$ on $M$. Dividing by edge length gives
\begin{equation}
\frac{\Delta_e y}{\Delta_e} = \langle \nabla_M y(x_v), \xi_e \rangle + O(\Delta_e).
\end{equation}

Similarly, for the second function,
\begin{equation}
\frac{\Delta_e z}{\Delta_e} = \langle \nabla_M z(x_v), \xi_e \rangle + O(\Delta_e).
\end{equation}

The numerator of $\text{cm}_{\text{cor},\partial}(y,z)(v)$ aggregates products
of these discrete directional derivatives over the neighborhood $N(v)$. As
$n \to \infty$ and the neighborhood becomes dense, this sum approximates a
Riemann integral over directions. Specifically, for uniformly distributed
neighbors in a geodesic ball, the directions $\{\xi_e : u \in N(v)\}$ become
equidistributed on the unit sphere $S^{d-1} \subset T_{x_v}M$ in the tangent
space. We obtain
\begin{align}
\sum_{u \in N(v)} \frac{\Delta_e y}{\Delta_e} \cdot \frac{\Delta_e z}{\Delta_e}
&\approx \sum_{u \in N(v)} \langle \nabla_M y(x_v), \xi_e \rangle \langle \nabla_M z(x_v), \xi_e \rangle \\
&\to |N(v)| \int_{S^{d-1}} \langle \nabla_M y(x_v), \xi \rangle \langle \nabla_M z(x_v), \xi \rangle \, d\sigma(\xi),
\end{align}
where $d\sigma$ denotes the uniform probability measure on the unit sphere
$S^{d-1}$.

We evaluate this integral using a standard result from directional statistics.
For any two vectors $\mathbf{a}, \mathbf{b} \in \mathbb{R}^d$, the integral of
their directional projections squared over the unit sphere satisfies
\begin{equation}
\int_{S^{d-1}} \langle \mathbf{a}, \xi \rangle \langle \mathbf{b}, \xi \rangle \, d\sigma(\xi) = \frac{1}{d} \langle \mathbf{a}, \mathbf{b} \rangle.
\end{equation}

This identity follows from expanding the inner products in coordinates and using
the fact that for $i \neq j$, we have
$\int_{S^{d-1}} \xi_i \xi_j \, d\sigma(\xi) = 0$ by symmetry, while
$\int_{S^{d-1}} \xi_i^2 \, d\sigma(\xi) = 1/d$ by normalization (since
$\sum_{i=1}^d \int_{S^{d-1}} \xi_i^2 \, d\sigma(\xi) = \int_{S^{d-1}} |\xi|^2 \, d\sigma(\xi) = 1$
and all $d$ components contribute equally). Applying this to
$\mathbf{a} = \nabla_M y(x_v)$ and $\mathbf{b} = \nabla_M z(x_v)$, we obtain
\begin{equation}
\sum_{u \in N(v)} \frac{\Delta_e y}{\Delta_e} \cdot \frac{\Delta_e z}{\Delta_e} \to \frac{|N(v)|}{d} \langle \nabla_M y(x_v), \nabla_M z(x_v) \rangle.
\end{equation}

The denominator terms undergo similar approximation. We have
\begin{align}
\sum_{u \in N(v)} \left(\frac{\Delta_e y}{\Delta_e}\right)^2
&\approx \sum_{u \in N(v)} \langle \nabla_M y(x_v), \xi_e \rangle^2 \\
&\to |N(v)| \int_{S^{d-1}} \langle \nabla_M y(x_v), \xi \rangle^2 \, d\sigma(\xi) \\
&= \frac{|N(v)|}{d} |\nabla_M y(x_v)|^2,
\end{align}
where the final equality again uses the directional integral formula with $\mathbf{a} = \mathbf{b} = \nabla_M y(x_v)$. Similarly,
\begin{equation}
\sum_{u \in N(v)} \left(\frac{\Delta_e z}{\Delta_e}\right)^2 \to \frac{|N(v)|}{d} |\nabla_M z(x_v)|^2.
\end{equation}

Combining these results, we obtain the limiting formula:
\begin{align}
\lim_{n \to \infty} \text{cm}_{\text{cor},\partial}(y,z)(v)
&= \frac{\frac{|N(v)|}{d} \langle \nabla_M y(x_v), \nabla_M z(x_v) \rangle}{\sqrt{\frac{|N(v)|}{d} |\nabla_M y(x_v)|^2} \sqrt{\frac{|N(v)|}{d} |\nabla_M z(x_v)|^2}} \\
&= \frac{\langle \nabla_M y(x_v), \nabla_M z(x_v) \rangle}{|\nabla_M y(x_v)| \cdot |\nabla_M z(x_v)|} \\
&= \cos \theta(x_v),
\end{align}
where $\theta(x_v) \in [0, \pi]$ denotes the angle between the gradient vectors
$\nabla_M y(x_v)$ and $\nabla_M z(x_v)$ in the tangent space $T_{x_v}M$.

This limiting interpretation provides compelling geometric meaning for the
correlation-type co-monotonicity coefficient. When
$\text{cm}_{\text{cor},\partial}(y,z)(v) \approx 1$, the gradients of $y$ and
$z$ are nearly parallel at the corresponding point in $M$, indicating that both
functions increase in approximately the same direction. When
$\text{cm}_{\text{cor},\partial}(y,z)(v) \approx -1$, the gradients are
anti-parallel, so the functions increase in opposite directions. When
$\text{cm}_{\text{cor},\partial}(y,z)(v) \approx 0$, the gradients are
approximately orthogonal, indicating that the functions vary independently in
different tangent directions.

This geometric perspective explains why derivative weighting is natural for
smooth functions on manifolds: it captures intrinsic geometric association that
is independent of the ambient coordinate system used to represent the data. The
normalization by edge length converts raw differences into discrete
approximations of directional derivatives, which are the natural geometric
quantities for characterizing function variation on Riemannian manifolds.

In finite samples with discrete neighborhoods, this limiting result suggests why
the correlation-type formulation performs well in practice. The asymptotic
formula assumes the neighborhood becomes dense in the tangent space, providing
approximately uniform directional coverage. With finite $k$, the discrete
neighborhood necessarily undersamples certain tangent directions, but the
correlation-type normalization remains stable provided both functions exhibit
sufficient variation across incident edges. The geometric mean in the denominator
ensures the coefficient scales appropriately even when gradient magnitudes vary
spatially, making the measure robust to heterogeneous signal patterns common in
high-dimensional biological data.

\subsection*{4.4 Edge Weighting Schemes}

The correlation-type co-monotonicity coefficient defined in Section 4.2 allows
flexible edge weighting through the parameters $w_e \geq 0$. The choice of
weights determines which aspects of directional concordance the measure
emphasizes, with different schemes appropriate for different data structures and
inferential goals. We examine three canonical weighting schemes that arise
naturally in geometric data analysis.

The simplest choice assigns uniform weight to all edges: $w_e = 1$ for all
$e \in E$. This unit weighting scheme treats edges equally regardless of their
length, emphasizing the combinatorial structure of the graph. The resulting
coefficient
\begin{equation}
\text{cm}_{\text{cor}}(y,z;1)(v) = \frac{\sum_{u \in N(v)} \Delta_e y \cdot \Delta_e z}{\sqrt{\sum_{u \in N(v)} (\Delta_e y)^2} \sqrt{\sum_{u \in N(v)} (\Delta_e z)^2}}
\end{equation}
directly correlates the raw function differences across edges without geometric
normalization. Unit weighting is appropriate when edge lengths are roughly
comparable, as in regular lattices or uniformly sampled grids, or when the graph
represents abstract relationships rather than geometric proximity. For graphs
constructed from irregular spatial samples where edge lengths vary substantially,
unit weighting may overweight contributions from long edges that span large
distances in the underlying space, obscuring local patterns of directional
concordance.

To account for geometric scaling, we employ derivative weighting as introduced in
Section 4.2. Setting $w_e = 1/\ell_e$ where $\ell_e$ denotes edge length yields
\begin{equation}
\text{cm}_{\text{cor},\partial}(y,z)(v) = \frac{\sum_{u \in N(v)} \frac{\Delta_e y}{\ell_e} \cdot \frac{\Delta_e z}{\ell_e}}{\sqrt{\sum_{u \in N(v)} \left(\frac{\Delta_e y}{\ell_e}\right)^2} \sqrt{\sum_{u \in N(v)} \left(\frac{\Delta_e z}{\ell_e}\right)^2}},
\end{equation}
which correlates discrete directional derivatives rather than raw differences.
This normalization by edge length has two beneficial effects. First, it prevents
long edges from dominating the coefficient: a large absolute change
$\Delta_e y$ across a long edge $\ell_e$ corresponds to a moderate derivative
$\Delta_e y / \ell_e$, giving it appropriate weight relative to shorter edges.
Second, as established in Section 4.3, derivative weighting ensures that the
coefficient approximates the cosine of the angle between gradient vectors on
smooth manifolds, providing intrinsic geometric meaning independent of
coordinate representation. Derivative weighting is appropriate when functions
represent continuous quantities sampled at irregular positions and we wish to
assess co-variation in the underlying continuous fields rather than artifacts of
the sampling pattern.

A third weighting scheme arises naturally from spectral graph theory and
diffusion processes on graphs. In the graph Laplacian framework, edge weights
often represent conductances $c_e$ that govern the rate of diffusion between
adjacent vertices. Using $w_e = c_e$ yields a coefficient that emphasizes edges
with high conductance, reflecting the effective connectivity structure rather
than purely geometric proximity. For graphs derived from kernel-weighted
similarity matrices where $c_e = \exp(-d_e^2/\sigma^2)$ with $d_e$ denoting
distance and $\sigma$ controlling bandwidth, conductance weighting downweights
distant connections while preserving strong local relationships. This scheme
proves useful when the graph encodes probabilistic or functional relationships
rather than strict geometric structure, as in biological networks where
interaction strength varies independently of physical distance.

The choice among these weighting schemes depends on the data structure and the
inferential question. Derivative weighting is generally preferred for spatially
embedded data where geometric scaling matters and functions are approximately
continuous. Unit weighting suffices for combinatorial graphs or when edge
lengths are comparable. Conductance weighting addresses situations where edge
importance varies beyond geometric considerations, capturing the effective
strength of relationships encoded by the graph structure. In all cases, the
correlation-type normalization ensures coefficients remain bounded in $[-1,1]$
and retain interpretability as measures of directional concordance scaled by the
geometric mean of function variations.

\subsection*{4.5 Scale Artifacts and Geometric Smoothing}

The unit and derivative weighting schemes measure directional concordance at different geometric scales, which can produce substantial discrepancies when graph edge lengths vary heterogeneously. We demonstrate this phenomenon through an empirical analysis of spontaneous preterm birth associations in vaginal microbiome data (Figure \ref{fig:cm_cor_comparison}), then introduce geometric smoothing as a principled approach for scale-consistent inference.

Consider a k-nearest neighbor graph constructed from compositional data, where edge lengths $\ell_e$ represent distances between samples in a high-dimensional feature space. The distribution of edge lengths typically exhibits considerable heterogeneity: most edges connect samples at moderate distances (say, $0.05 < \ell_e < 0.15$), but some edges span very short distances ($\ell_e < 0.01$) between nearly identical samples arising from repeated measurements, high sampling density in certain regions, or genuine biological clustering. Under derivative weighting with $w_e = 1/\ell_e$, these very short edges receive extreme weights that can exceed typical edge weights by factors of $10^3$ to $10^5$.

We assessed the practical impact of scale heterogeneity by computing both unit-weighted and derivative-weighted correlation coefficients between sPTB prevalence and 106 phylotype abundances across 2,117 vertices in a vaginal microbiome graph, yielding 224,202 vertex-phylotype pairs. Figure \ref{fig:cm_cor_comparison}A displays the relationship between these two measures for all pairs. While most pairs cluster near the diagonal, indicating agreement between weighting schemes, substantial discrepancies appear at 964 pairs (0.43\% of total, highlighted in red) where $|\text{cm}_{\text{cor}}(y,z;1)(v) - \text{cm}_{\text{cor},\partial}(y,z)(v)| > 0.75$. The scatter reveals that unit and derivative weighting can yield markedly different assessments of the same biological association. One extreme case displayed opposite signs: unit weighting yielded $-0.618$ while derivative weighting yielded $+0.906$ for the association between sPTB and Corynebacterium imitans at a specific vertex.

Detailed examination of this extreme case revealed the mechanism underlying such discrepancies. The vertex had 37 neighbors, but five edges with lengths below 0.005 received derivative weights exceeding 200-fold typical values, dominating the calculation and yielding opposite-sign concordance from the broader neighborhood pattern. The unit-weighted coefficient averaged uniformly across all 37 edges, yielding a negative association reflecting the majority pattern. The derivative-weighted coefficient was driven almost entirely by the five extremely short edges, which exhibited positive concordance while the remaining 32 edges showed negative concordance.

\begin{center}
\includegraphics[width=0.48\textwidth]{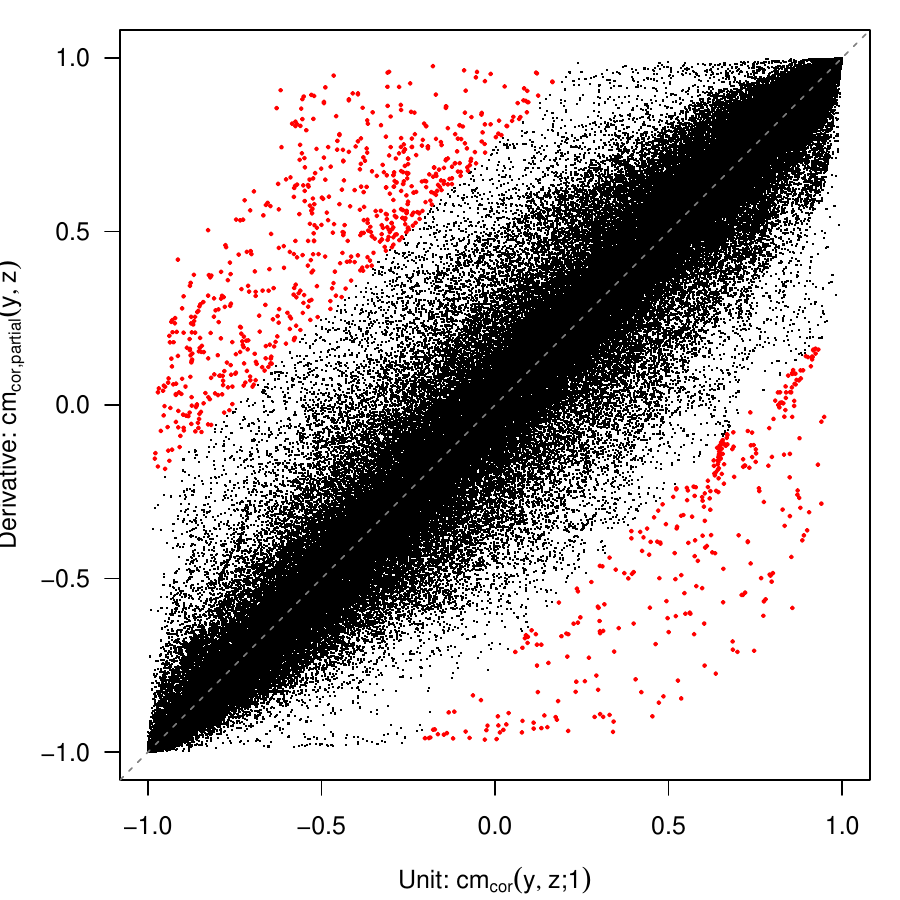}
\includegraphics[width=0.48\textwidth]{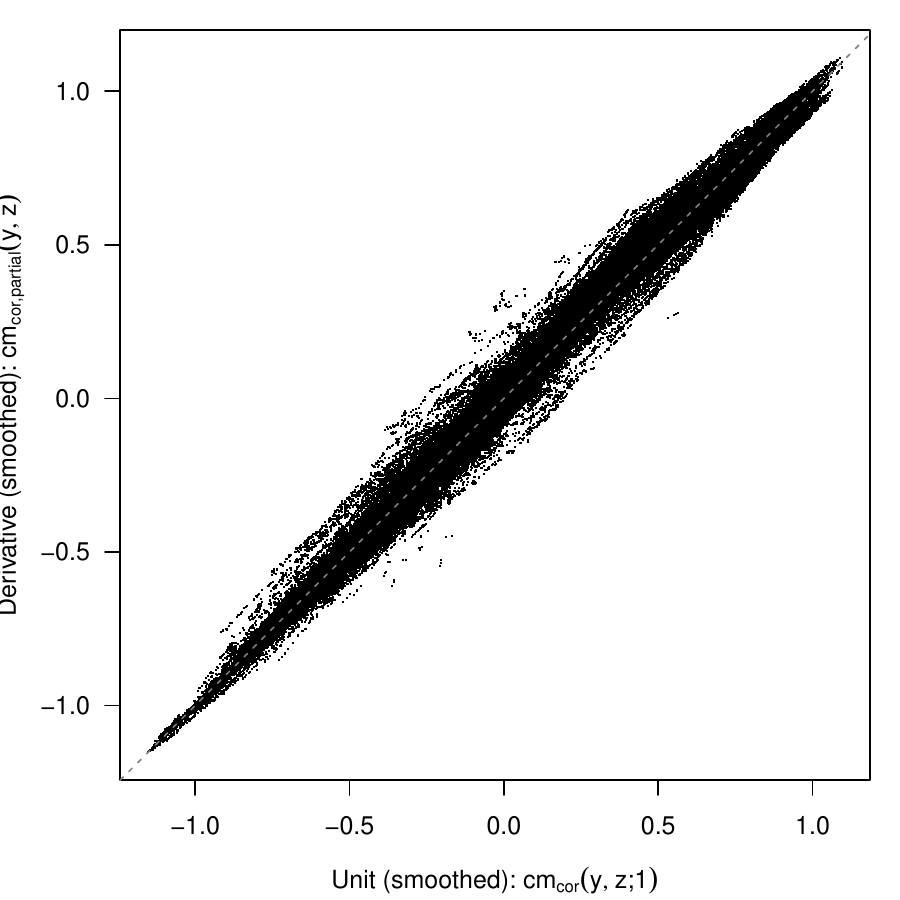}
\captionof{figure}{Scale artifacts in co-monotonicity coefficients are eliminated by geometric smoothing. (A) Raw correlation-type coefficients computed with derivative versus unit weighting for sPTB prevalence and 106 phylotypes across 224,402 vertex-phylotype pairs. Red points indicate pairs with large discrepancies ($|\text{difference}| > 0.75$). (B) The same coefficients after geometric smoothing via graph Laplacian filtering. Smoothing eliminates scale-dependent discrepancies, demonstrating that both weighting schemes recover nearly identical association structure at consistent geometric scale.}
\label{fig:cm_cor_comparison}
\end{center}

This behavior reflects fundamentally different notions of local association. Unit weighting measures concordance across the combinatorial neighborhood, treating all connected vertices equally regardless of their geometric separation. Derivative weighting measures concordance in the geometric tangent space, where very short edges approximate infinitesimal neighborhoods and receive proportionally large weight according to the distance gradient. When micro-scale patterns at edges shorter than 0.01 differ from meso-scale patterns at edges between 0.05 and 0.15, the two measures capture genuinely distinct geometric phenomena operating at different scales. The derivative-weighted coefficient becomes sensitive to association structure at arbitrarily fine scales determined by the shortest edges in the graph, which may reflect sampling artifacts rather than meaningful biological variation. Conversely, the unit-weighted coefficient treats all scales equally within the combinatorial neighborhood, potentially obscuring geometric structure encoded in edge lengths. Without additional constraints, neither measure uniquely captures the associations of scientific interest.

These findings demonstrate that raw co-monotonicity coefficients exhibit scale-dependent artifacts that obscure biological signal. We address this scale heterogeneity through geometric smoothing of the computed coefficient matrices. The graph Laplacian low-pass filter employed for smoothing response functions (Section 3) naturally extends to smoothing association maps. For a co-monotonicity matrix $\text{CM}(y, Z)$ with dimensions $n \times m$, we apply the smoothing operation to each column independently:
\begin{equation}
\text{CM}_{\text{smooth}}(y, Z) = [\text{smooth}(\text{CM}(y, Z)[:,j])]_{j=1,\ldots,m},
\end{equation}
where the smoothing uses the same regularization parameter $\lambda$ selected for the response $y$ via generalized cross-validation or similar criteria. This ensures that associations are analyzed at the same geometric scale as the response variation itself, maintaining consistency throughout the analysis pipeline.

Figure \ref{fig:cm_cor_comparison}B displays the result of applying this smoothing procedure to both unit-weighted and derivative-weighted coefficients. The effect is substantial: the mean absolute difference between the two measures drops from 0.089 to 0.029, the 95th percentile decreases from 0.329 to 0.095, and all 964 pairs with raw differences exceeding 0.75 reduce to disagreements below this threshold after smoothing. The smoothed coefficients align almost perfectly along the diagonal, with correlation exceeding 0.999 between the two measures. The extreme case of opposite signs resolved to agreement: both smoothed coefficients yielded $+0.82$, confirming a positive association at the biologically relevant scale while filtering out the micro-scale artifact from very short edges. This near-perfect concordance after smoothing demonstrates that unit and derivative weighting recover the same underlying association structure once analyzed at a consistent geometric scale.

\begin{center}
\includegraphics[width=0.6\textwidth]{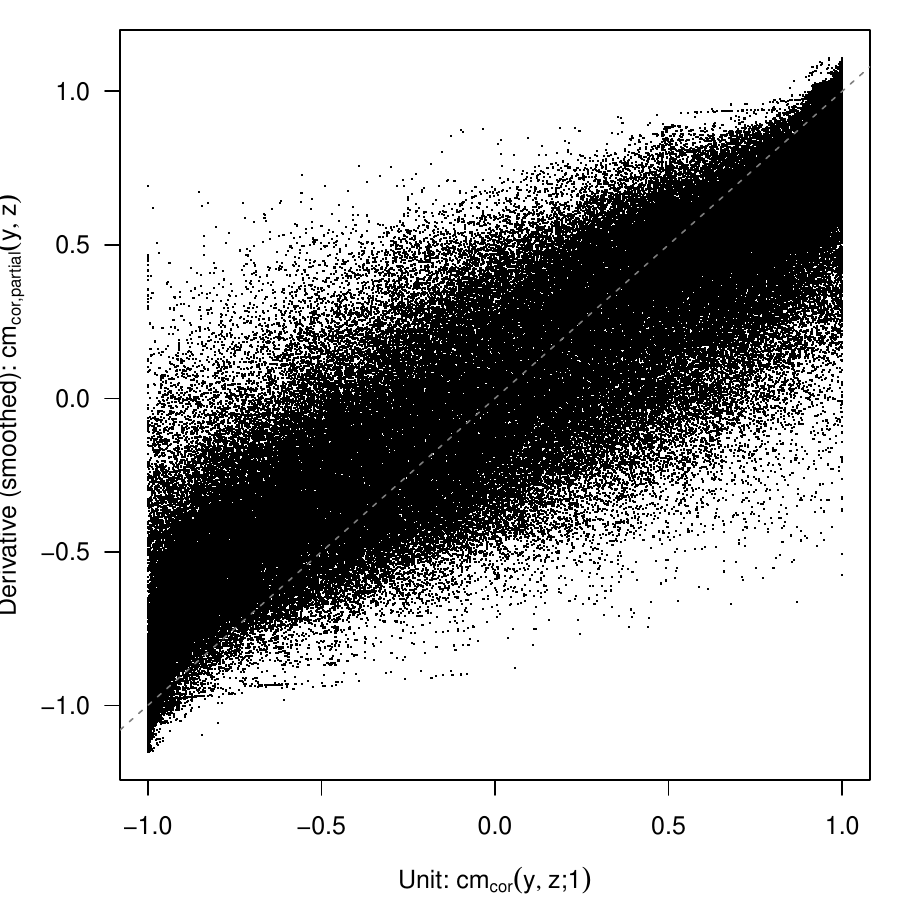}
\captionof{figure}{Raw unit-weighted coefficients do not match smoothed derivative-weighted coefficients, demonstrating that smoothing fundamentally transforms both weighting schemes. Comparison of raw unit-weighted co-monotonicity coefficients versus smoothed derivative-weighted coefficients for sPTB prevalence and 106 phylotypes across 224,402 vertex-phylotype pairs. The wide scatter (Gini mean difference: 0.322) contrasts with the tight agreement between smoothed unit-weighted and smoothed derivative-weighted coefficients (Figure~\ref{fig:cm_cor_comparison}B), confirming that geometric smoothing modifies association structure rather than merely adjusting one scheme to match the other.}
\label{fig:raw_unit_smoothed_deriv}
\end{center}

Importantly, both unit-weighted and derivative-weighted coefficients are substantially modified by the smoothing process. Figure~\ref{fig:raw_unit_smoothed_deriv} compares raw unit-weighted coefficients with smoothed derivative-weighted coefficients, revealing persistent discrepancies throughout the distribution. While the mean difference is near zero ($-0.007$), substantial scatter remains with the 5th and 95th percentiles at $-0.490$ and $0.453$ respectively, range from $-1.69$ to $1.57$, and Gini mean difference of $0.322$. This wide scatter contrasts sharply with the tight agreement between smoothed unit-weighted and smoothed derivative-weighted coefficients (Figure~\ref{fig:cm_cor_comparison}B), confirming that geometric smoothing substantively modifies the association structure for both weighting schemes rather than simply adjusting one to match the raw values of the other.

This empirical validation reveals that the apparent conflict between unit and derivative weighting reflects scale artifacts rather than fundamental methodological differences. The biologically meaningful associations exist at the geometric scale determined by the graph Laplacian regularization, which balances signal fidelity against noise suppression. Micro-scale variations captured by extreme derivative weights on very short edges represent either sampling artifacts such as repeated measurements and density fluctuations, or true but scientifically irrelevant local fluctuations. The smoothing operation removes precisely these high-frequency components, leaving the large-scale association structure that both normalizations agree upon. This convergence at consistent geometric scales indicates that the choice of weighting scheme becomes largely immaterial after appropriate smoothing. Both smoothed unit-weighted and smoothed derivative-weighted coefficients provide equally defensible measures of association structure. We adopt smoothed derivative-weighted coefficients for subsequent analyses, as derivative weighting naturally adapts to local manifold geometry, but we acknowledge this choice is pragmatic rather than definitive.

\subsection*{4.6 Matrix Extension for Multivariate Analysis}

In practice, we rarely analyze a single predictor in isolation. Microbiome
studies measure hundreds of bacterial taxa, genomic studies measure thousands of
genes, and network data involve numerous node attributes. We require efficient
computation of co-monotonicity between an outcome and multiple features, as well
as among features themselves.

Let $y: V \to \mathbb{R}$ be an outcome function and
$Z = [z_1, \ldots, z_m]: V \to \mathbb{R}^m$ be a matrix of $m$ feature
functions. The co-monotonicity matrix between $y$ and $Z$ is the $n \times m$
matrix
\begin{equation}
\text{CM}(y, Z) = [\text{cm}(y, z_j;w)(v)]_{v \in V, j=1,\ldots,m},
\end{equation}
where each column $j$ contains the vertex-wise co-monotonicity coefficients
between $y$ and feature $z_j$. This matrix provides association profiles: row
$v$ shows how all features associate with the outcome in the neighborhood of
vertex $v$, while column $j$ shows the spatial pattern of association for
feature $z_j$ across all vertices.

Computing $\text{CM}(y, Z)$ naively by $m$ independent calls to the pairwise
co-monotonicity function would redundantly recompute edge differences
$\Delta_e y$ for each feature. We optimize by precomputing these $y$-dependent
quantities once, then reusing them across all features. The algorithm maintains
edge weights $w_e$ and differences $\Delta_e y$ in memory, iterates over
features $z_j$, and for each feature computes $\Delta_e z_j$ and accumulates
weighted products. This reduces computational cost from $O(m \cdot |E|)$
operations with independent overhead to $O(|E| + m \cdot |E|)$ with shared
preprocessing, yielding substantial savings when $m$ is large.

For feature-feature associations, we compute the tensor
\begin{equation}
\text{CM}(Z, Z) = [\text{cm}(z_j, z_k;w)(v)]_{v \in V, j,k=1,\ldots,m},
\end{equation}
which is an $n \times m \times m$ object containing co-monotonicity between all
pairs of features at all vertices. The slice $\text{CM}(Z, Z)[v, :, :]$ is an
$m \times m$ matrix showing how features co-vary in the neighborhood of vertex
$v$. This captures local correlation structure in the feature space, revealing
whether features cluster into modules that vary coherently or whether feature
relationships differ across regions of the graph.

In applications, we augment $\text{CM}(y, Z)$ with selected columns from
$\text{CM}(Z, Z)$ to form an extended association profile. For a vertex $v$,
this profile includes both $\text{cm}(y, z_j)(v)$ for all features $j$ and
$\text{cm}(z_j, z_k)(v)$ for pairs $(j,k)$ of interest. These profiles embed
vertices into an association space where proximity reflects similarity in how
outcome and features relate locally. Vertices with similar profiles form
co-monotonicity cells, which we will explore in Section 5 as a basis
for geometric multiple testing and biclustering.

\subsection*{4.7 Alternative Normalizations}

The correlation-type co-monotonicity coefficient developed in Sections 4.2-4.4
employs Pearson-style normalization, dividing by the geometric mean of squared
variations. This normalization provides stable estimates and interpretable
coefficients bounded in $[-1,1]$ for most applications. However, specific data
characteristics or inferential goals may suggest alternative normalization
schemes. We examine three such alternatives, providing concrete examples that
illustrate when each approach succeeds or fails.

We begin by considering absolute value normalization, which divides the sum of
weighted products by the sum of weighted absolute products. For edge weights
$w_e$, this yields
\begin{equation}
\text{cm}_{\text{abs}}(y,z;w)(v) = \frac{\sum_{u \in N(v)} w_e \Delta_e y \cdot \Delta_e z}{\sum_{u \in N(v)} w_e |\Delta_e y \cdot \Delta_e z|}.
\end{equation}
This normalization has intuitive interpretation: it measures the balance between
concordant edges (positive products) and discordant edges (negative products),
weighted by the magnitude of their products. When all edges show agreement, the
numerator equals the denominator and $\text{cm}_{\text{abs}} = 1$. When
disagreements exactly balance agreements in product-weighted terms, the
coefficient vanishes.

The critical weakness of absolute value normalization emerges when one function
varies sparsely. Consider a vertex $v$ with 30 neighbors where the outcome $y$
exhibits variation of magnitude 0.1 across all edges, while feature $z$ shows
meaningful variation (magnitude 0.1) on only one edge and remains essentially
constant (changes $< 0.01$) on the remaining 29 edges. The single edge with
signal contributes product $\pm 0.01$ to the numerator. The 29 near-constant
edges contribute products near zero. The denominator sums the absolute values of
these products, yielding approximately $0.01 + 29 \times 0.001 = 0.039$. The
resulting coefficient is $\pm 0.01 / 0.039 \approx \pm 0.26$, suggesting
moderate association despite signal appearing on only $1/30 \approx 3\%$ of
edges. As the non-signal edges become exactly constant, the coefficient
approaches $\pm 1$, indicating perfect association based on a single edge. This
instability arises because the denominator can become arbitrarily small when one
function is nearly constant, allowing sparse signal to dominate the measure.

Sign-based normalization addresses magnitude sensitivity by considering only
directional information. We replace the weighted sums with counts of edges
showing agreement versus disagreement:
\begin{equation}
\text{cm}_{\text{sign}}(y,z)(v) = \frac{\sum_{u \in N(v)} \text{sign}(\Delta_e y \cdot \Delta_e z)}{|N(v)|},
\end{equation}
where $\text{sign}(x) = 1$ if $x > 0$, $\text{sign}(x) = -1$ if $x < 0$, and
$\text{sign}(0) = 0$. This coefficient measures the proportion of edges where
functions agree in direction minus the proportion where they disagree, treating
all edges equally regardless of change magnitude.

Sign-based normalization discards potentially valuable information about the
strength of associations. Consider two scenarios at a vertex with 20 neighbors.
In the first scenario, $y$ and $z$ both change by 0.5 units in the same
direction on 11 edges and in opposite directions on 9 edges, yielding
substantial co-variation with correlation 0.7. In the second scenario, $y$
changes by 0.5 units on all edges while $z$ shows tiny changes of 0.01 units in
directions that happen to align with $y$ on 11 edges and oppose $y$ on 9 edges,
exhibiting negligible true co-variation. The sign-based coefficient produces
$(11-9)/20 = 0.1$ in both cases, treating large coordinated changes identically
to small random fluctuations. The magnitude information that distinguishes
biologically meaningful co-variation from numerical noise is completely lost.
Moreover, sign-based coefficients remain vulnerable to the sparse signal problem:
when $z$ varies on only one edge, that single edge's sign determines the
coefficient, yielding values of $\pm 1/|N(v)|$ that overstate association
strength relative to the prevalence of signal.

These limitations of absolute value and sign-based normalizations arise from
their treatment of edges with weak or absent signal. The absolute value
denominator becomes small when products are small, amplifying sparse signals. The
sign-based approach counts edges with arbitrarily weak variation equally with
edges showing strong variation, obscuring signal strength. In practice, neither
normalization provides reliable inference for heterogeneous data where features
exhibit spatially varying signal intensity. We do not recommend their use in
applications requiring robust, interpretable association measures.

For scenarios involving sparse or binary features where explicit signal filtering
is scientifically justified, we introduce proportion-based co-monotonicity with
thresholds. This measure explicitly filters edges by signal strength before
aggregating directional agreements. Given threshold parameters $\tau_y, \tau_z > 0$,
we assign each edge $e = [v,u]$ a directional agreement score:
\begin{equation}
s_e = \begin{cases}
+1 & \text{if } |\Delta_e y| > \tau_y, |\Delta_e z| > \tau_z, \text{ and } \Delta_e y \cdot \Delta_e z > 0 \\
-1 & \text{if } |\Delta_e y| > \tau_y, |\Delta_e z| > \tau_z, \text{ and } \Delta_e y \cdot \Delta_e z < 0 \\
0 & \text{otherwise.}
\end{cases}
\end{equation}
The proportion-based coefficient at vertex $v$ is then
\begin{equation}
\text{cm}_{\text{prop}}(y,z;\tau_y,\tau_z)(v) = \frac{1}{|N(v)|}\sum_{u \in N(v)} s_e,
\end{equation}
where the sum is over all edges incident to $v$, and $|N(v)|$ denotes the total
number of neighbors. Edges where either function shows insufficient variation
(below threshold) contribute zero rather than being excluded from the
denominator.

The key distinction from previous measures lies in how sparse signal is handled.
Return to the vertex with 30 neighbors where $y$ varies on all edges but $z$
shows meaningful signal on only one edge. Setting thresholds $\tau_y = \tau_z = 0.05$
appropriate for the variation scales, the proportion-based measure yields
$\text{cm}_{\text{prop}}(y,z)(v) = \pm 1/30 \approx \pm 0.033$. This value
correctly indicates that association is supported by sparse evidence: only $3\%$
of the neighborhood exhibits coordinated variation. As more edges develop signal,
the coefficient magnitude increases proportionally, reaching $\pm 1$ only when
directional concordance appears across the entire neighborhood. The fixed
denominator prevents sparse signals from producing misleadingly large
coefficients while maintaining interpretability as the proportion of edges
showing directional agreement.

Threshold selection depends on data characteristics and inferential goals. For
the response function $y$, thresholds can be set relative to overall variation:
$\tau_y = c \cdot \text{sd}(y)$ for small $c$ (typically $0.05$ to $0.10$)
filters changes smaller than $5-10\%$ of a standard deviation, removing
numerical noise while preserving meaningful signal. For feature $z$ representing
quantities like phylotype abundances that vary heterogeneously across features,
adaptive per-feature thresholds based on quantiles of edge differences prove
effective. Setting $\tau_{z_j} = Q_{0.25}(\{|\Delta_e z_j| : e \in E\})$ as the
first quartile of absolute edge differences for feature $z_j$ filters the bottom
$25\%$ of changes as noise while preserving variation in the upper three
quartiles. This adaptive approach handles features with different baseline
variability without requiring manual tuning. When thresholds are set to zero
($\tau_y = \tau_z = 0$), the proportion-based measure reduces to the sign-based
coefficient, recovering pure directional counting without magnitude filtering.

The proportion-based measure is particularly appropriate for binary or sparse
features where the presence or absence of variation carries biological meaning.
In microbiome studies, phylotypes may be absent from most samples but abundant
where present, exhibiting extreme sparsity. The proportion-based coefficient
with appropriate thresholds identifies regions where a phylotype's presence
correlates with outcome changes while appropriately penalizing associations
supported by few samples. For continuous features exhibiting widespread
variation, the correlation-type measure from Section 4.2 generally provides more
sensitive detection of graded co-variation by weighting edges according to the
magnitude of coordinated changes.

We summarize the comparison between normalization schemes. The correlation-type
measure $\text{cm}_{\text{cor}}$ employs geometric mean normalization that
provides stable coefficients across diverse signal patterns, interprets
naturally as correlation between discrete derivatives, and weights edges by the
magnitude of coordinated variation. It is recommended for general use with
continuous features exhibiting heterogeneous but widespread variation. The
proportion-based measure $\text{cm}_{\text{prop}}$ with thresholds explicitly
filters weak signals, interprets as the proportion of neighborhood edges showing
directional agreement, and naturally handles sparse or binary features where
signal prevalence matters. It is recommended when explicit signal filtering is
scientifically justified or when features exhibit extreme sparsity. The absolute
value normalization $\text{cm}_{\text{abs}}$ becomes unstable with sparse
signals and is not recommended for practical use. The sign-based normalization
$\text{cm}_{\text{sign}}$ discards magnitude information that typically proves
essential for distinguishing meaningful associations from noise and is not
recommended for practical use. In applications where the choice is unclear, we
suggest computing both $\text{cm}_{\text{cor}}$ and $\text{cm}_{\text{prop}}$
with conservative thresholds, examining their agreement to assess whether
magnitude weighting versus prevalence counting yields substantively different
conclusions.

\section*{5. Statistical Inference for Co-Monotonicity}

Having developed co-monotonicity coefficients as geometric measures of directional
association, we confront the fundamental inferential question: how should we distinguish
genuine associations from artifacts of sampling variability? The vertex-wise nature of
co-monotonicity coefficients presents both opportunity and challenge. We obtain spatially
resolved information revealing where associations are strongest, but this spatial
resolution demands vertex-level inference procedures that respect the graph structure
while accounting for multiple comparisons. The standard hypothesis testing framework,
with its focus on $p$-values and frequentist error rates, sits uncomfortably with the
Bayesian spectral filtering that produces smoothed function estimates. We require
inference procedures that propagate uncertainty from the estimation stage through to
association quantification, providing probabilistic statements about effect sizes rather
than merely rejecting null hypotheses.

\subsection*{5.1 The Multiple Testing Challenge}

Consider the typical scenario in microbiome-outcome association studies. We have $n$
samples (vertices in the graph $G$), an outcome $y: V \to \mathbb{R}$, and $m$ bacterial
taxa (features) $Z = [z_1, \ldots, z_m]: V \to \mathbb{R}^m$. After smoothing to obtain
$\hat{y}$ and $\hat{Z}$ through spectral filtering, we compute the co-monotonicity
matrix $\text{CM}(\hat{y}, \hat{Z}) \in \mathbb{R}^{n \times m}$ with
$n \times m$ coefficients. A naive testing procedure would assess each coefficient
independently: for each vertex-feature pair $(v, j)$, test whether
$\text{cm}(\hat{y}, z_j)(v)$ differs significantly from zero. This generates $nm$
hypothesis tests, and with typical values $n = 200$ samples and $m = 100$ features, we
face $20{,}000$ simultaneous tests.

The multiple testing burden is severe. Even with independent tests, controlling the
family-wise error rate at level $\alpha = 0.05$ through Bonferroni correction requires
declaring significance only when $p < 0.05/20000 = 2.5 \times 10^{-6}$, a threshold so
stringent that only the most extreme associations would be detected. False discovery
rate control offers less conservative correction but still penalizes for the sheer
number of tests. Moreover, the independence assumption underlying standard FDR
procedures fails dramatically in our setting: co-monotonicity coefficients at adjacent
vertices are inherently correlated because they aggregate over overlapping
neighborhoods. The graph structure induces spatial dependence that standard multiple
testing corrections ignore.

Beyond the computational burden of multiple testing, there is a conceptual issue. The
classical hypothesis testing framework asks whether we can reject the null hypothesis of
no association, yielding binary decisions (reject or fail to reject) with
probabilistic error control. Yet in exploratory high-dimensional data analysis, we
rarely seek such binary classifications. We wish to identify regions where associations
are strongest, quantify the magnitude of these associations with uncertainty bounds, and
compare the strength of different feature-outcome relationships. The hypothesis testing
paradigm, with its focus on $p$-values that measure tail probabilities under null
distributions, provides limited information for these goals. We require inference
methods that directly address probabilistic questions about effect sizes: what is
$P(|\text{cm}(\hat{y}, z_j)(v)| > \delta \mid \text{data})$ for scientifically
meaningful thresholds $\delta$, rather than merely $P(\text{data} \mid \text{cm} = 0)$?

\subsection*{5.2 Vertex-Wise Permutation Testing}

We begin with the classical nonparametric approach to association testing. Permutation
tests provide exact finite-sample inference without distributional assumptions, making
them particularly appealing for complex data structures. The key insight is that under
the null hypothesis of no association between feature $z$ and outcome $y$, permuting the
feature values across vertices should preserve all aspects of the data generation
process except the association of interest.

At each vertex $v$, we test the null hypothesis $H_0^v: \text{cm}(y, z)(v) = 0$, which
corresponds to asserting that $y$ and $z$ have no directional relationship in the
neighborhood of $v$. We compute the observed coefficient
$c_{\text{obs}}(v) = \text{cm}(\hat{y}, \hat{z})(v)$ using the smoothed estimates. To
generate the null distribution, we permute the feature values: let $\pi$ be a random
permutation of $\{1, \ldots, n\}$, and define the permuted feature
$z^{\pi}(v_i) = z(v_{\pi(i)})$ for $i = 1, \ldots, n$. This permutation breaks any
association between $z$ and $y$ while preserving the marginal distribution of $z$ and
the graph structure $G$.

For each permutation $\pi_b$ with $b = 1, \ldots, B$, we smooth the permuted feature to
obtain $\hat{z}^{(\pi_b)}$ and compute $c_b(v) = \text{cm}(\hat{y}, \hat{z}^{(\pi_b)})(v)$.
The collection $\{c_1(v), \ldots, c_B(v)\}$ forms an empirical null distribution for the
co-monotonicity coefficient at vertex $v$ under the hypothesis of no association. We
compute the two-sided $p$-value as
\begin{equation}
p(v) = \frac{1 + \#\{b : |c_b(v)| \geq |c_{\text{obs}}(v)|\}}{B + 1},
\end{equation}
where the numerator counts permutations yielding coefficients at least as extreme as the
observed value, and the denominator includes the observed data itself as one possible
permutation.

This vertex-wise permutation procedure respects the graph structure: the permuted
features are smoothed using the same spectral filtering as the original data, ensuring
that the null distribution reflects the spatial autocorrelation induced by the Laplacian.
However, the procedure treats the smoothed outcome $\hat{y}$ as fixed, ignoring the
uncertainty in its estimation. When $y$ itself is estimated through spectral filtering of
noisy observations, this omission can lead to underestimation of uncertainty and
inflated false positive rates.

For multiple testing correction across the $n$ vertices, we employ the Benjamini-Hochberg
procedure to control the false discovery rate. Order the $p$-values as
$p_{(1)} \leq p_{(2)} \leq \cdots \leq p_{(n)}$, and let $k^*$ be the largest $k$ such
that $p_{(k)} \leq (k/n) \alpha$ for target FDR level $\alpha$. We declare vertices
$v$ with $p(v) \leq p_{(k^*)}$ as significant. This procedure controls the expected
proportion of false discoveries among the rejected hypotheses, providing a less
conservative alternative to family-wise error rate control.

The spatial structure of significant vertices provides additional information. Rather
than treating each vertex independently, we identify connected components in the
subgraph induced by significant vertices. Isolated significant vertices likely represent
false positives that happened to achieve small $p$-values by chance, while spatially
coherent clusters of significant vertices suggest genuine regional associations. We can
impose a minimum cluster size threshold, declaring only those significant vertices
belonging to clusters of at least $k_{\min}$ vertices as discoveries. This spatial
thresholding provides robustness against false positives arising from the multiple
testing burden.

\subsection*{5.3 Bayesian Inference via Posterior Sampling}

The permutation testing framework, despite its nonparametric appeal, inherits the
limitations of the classical hypothesis testing paradigm. We now develop an alternative
Bayesian approach that addresses these limitations by treating uncertainty in the
smoothed estimates as the primary source of inferential variability.

The spectral filtering procedure that produces smoothed estimates $\hat{y}$ and $\hat{Z}$
from observed data $y$ and $Z$ admits natural Bayesian interpretation. Recall that the
filtered estimate takes the form $\hat{y} = V F_{\eta}(\Lambda) V^T y$, where $V$
contains eigenvectors of the normalized Laplacian, $\Lambda$ is the diagonal matrix of
eigenvalues, and $F_{\eta}$ is a spectral filter (such as the heat kernel
$F_{\eta}(\lambda) = \exp(-\eta \lambda)$ or Tikhonov filter
$F_{\eta}(\lambda) = 1/(1 + \eta \lambda)$). This operation corresponds to the posterior
mean under a Gaussian prior on spectral coefficients with precision proportional to
eigenvalues.

We elaborate this Bayesian perspective through a generative model for the observed data.
Suppose the true outcome function $f: V \to \mathbb{R}$ has spectral representation
$f = V \alpha$ for coefficients $\alpha \in \mathbb{R}^m$, where $m$ is the number of
eigenvectors retained. The smoothness prior on spectral coefficients is
\begin{equation}
p(\alpha \mid \eta) \propto \exp\left(-\frac{\eta}{2} \sum_{j=1}^m \lambda_j \alpha_j^2\right),
\end{equation}
which penalizes high-frequency modes (large $\lambda_j$) more severely. Given noisy
observations $y_v = f(v) + \epsilon_v$ with $\epsilon_v \sim N(0, \sigma^2)$, the
posterior distribution over spectral coefficients is Gaussian:
\begin{equation}
\alpha \mid y, \eta \sim N\left((I + \eta \Lambda)^{-1} V^T y,
\sigma^2 (I + \eta \Lambda)^{-1}\right).
\end{equation}

The filtered estimate $\hat{y} = V F_{\eta}(\Lambda) V^T y$ computes the posterior mean
when $F_{\eta}(\lambda) = 1/(1 + \eta \lambda)$. For other filter types, the filtered
estimate approximates the posterior mode or mean under related priors. This connection
suggests a natural approach to uncertainty quantification: sample from the posterior
distribution over spectral coefficients, transform to vertex space, and propagate this
uncertainty through the co-monotonicity computation.

However, directly sampling from the spectral coefficient posterior requires estimating
the residual variance $\sigma^2$ and assumes Gaussian noise, which may not hold for
discrete or bounded outcomes. We employ an alternative resampling strategy that induces
posterior-like variability without requiring parametric assumptions. The key idea is to
perturb the vertex masses in the Riemannian structure through Dirichlet resampling,
generating multiple weighted Laplacians that yield different smoothed estimates.

\begin{definition}[Dirichlet Resampling for Posterior Uncertainty]
Let $M_0$ be the vertex mass matrix with diagonal entries $M_0[v,v] = \rho_0(v)$. Denote
$w = (\rho_0(1)/n, \ldots, \rho_0(n)/n)$ as the normalized vertex weights summing to
unity. For concentration parameter $\alpha > 0$, we sample
\begin{equation}
w^* \sim \text{Dirichlet}(\alpha w_1, \ldots, \alpha w_n),
\end{equation}
where larger $\alpha$ concentrates the resampled weights near the original weights $w$,
and smaller $\alpha$ allows greater variability. We construct a perturbed mass matrix
$M_0^*$ with diagonal entries $M_0^*[v,v] = n w^*_v$, build the corresponding weighted
Laplacian $L^*$, and compute filtered estimates
$\hat{y}^* = V^* F_{\eta}(\Lambda^*) (V^*)^T y$ and
$\hat{z}_j^* = V^* F_{\eta}(\Lambda^*) (V^*)^T z_j$, where $V^*$ and $\Lambda^*$ come
from the eigendecomposition of $L^*$.
\end{definition}

This Dirichlet resampling procedure induces variability in the smoothed estimates through
perturbation of the geometric structure itself. Vertices with large mass $\rho_0(v)$ tend
to receive large perturbed mass $M_0^*[v,v]$, but stochastic variation allows for
reordering of vertex importance across samples. The resulting ensemble
$\{(\hat{y}^{(b)}, \hat{Z}^{(b)})\}_{b=1}^B$ for $B$ independent Dirichlet samples
approximates a posterior distribution over smoothed functions.

For each posterior sample $b$, we compute co-monotonicity coefficients
$c^{(b)}(v) = \text{cm}(\hat{y}^{(b)}, \hat{z}_j^{(b)})(v)$ at each vertex $v$ and for
each feature $j$. The empirical distribution of $\{c^{(1)}(v), \ldots, c^{(B)}(v)\}$
represents the posterior distribution of the co-monotonicity coefficient at vertex $v$.
We construct credible intervals by computing empirical quantiles: the
$(1-\alpha) \times 100\%$ credible interval for $\text{cm}(\hat{y}, z_j)(v)$ is
\begin{equation}
\text{CI}_{1-\alpha}(v) = [q_{\alpha/2}(\{c^{(b)}(v)\}),
q_{1-\alpha/2}(\{c^{(b)}(v)\})],
\end{equation}
where $q_p$ denotes the $p$-quantile of the empirical distribution.

These Bayesian credible intervals admit direct probability interpretation: given the
observed data and the smoothness prior implicit in spectral filtering, there is
$(1-\alpha) \times 100\%$ posterior probability that the true co-monotonicity lies within
the interval. This contrasts with frequentist confidence intervals, which make statements
about long-run coverage under repeated sampling. For a vertex $v$ where the credible
interval excludes zero (say, $\text{CI}_{0.95}(v) = [0.42, 0.71]$), we conclude with
high posterior probability that a genuine directional association exists in the
neighborhood of $v$, and the interval quantifies the magnitude of this association.

\subsection*{5.4 Posterior Probabilities for Effect Size Thresholds}

Beyond credible intervals, the Bayesian framework enables computation of posterior
probabilities for scientifically meaningful thresholds. Rather than testing whether
$\text{cm}(\hat{y}, z)(v) = 0$, we ask whether the co-monotonicity exceeds a threshold
$\delta > 0$ representing meaningful association strength. For example, in microbiome
studies, we might consider $|\text{cm}| > 0.3$ as indicating moderate directional
concordance, and $|\text{cm}| > 0.6$ as strong concordance.

The posterior probability that the co-monotonicity at vertex $v$ exceeds threshold
$\delta$ is estimated from the empirical distribution of posterior samples:
\begin{equation}
P(|\text{cm}(\hat{y}, z)(v)| > \delta \mid \text{data})
\approx \frac{1}{B} \#\{b : |c^{(b)}(v)| > \delta\}.
\end{equation}
This probability directly quantifies our belief that the association is meaningful,
given the data and prior assumptions. A vertex with
$P(|\text{cm}| > 0.3 \mid \text{data}) = 0.95$ has high posterior support for at least
moderate association, while $P(|\text{cm}| > 0.6 \mid \text{data}) = 0.45$ indicates
uncertain evidence for strong association.

This formulation avoids the binary decision problem inherent in hypothesis testing. We
need not declare associations as significant or non-significant based on arbitrary
$\alpha$ levels; instead, we report posterior probabilities and credible intervals,
allowing domain experts to interpret the strength of evidence in context. Moreover, the
Bayesian framework naturally handles multiple comparisons without explicit correction
procedures. The posterior distribution already accounts for all sources of uncertainty,
including estimation variability and the fact that we are simultaneously assessing many
vertex-feature pairs. There is no multiplicity penalty in the Bayesian paradigm because
we report probabilistic statements about effect sizes rather than making decisions about
null hypotheses.

\subsection*{5.5 Comparison of Inference Approaches}

The permutation testing and Bayesian posterior sampling approaches address the inferential
challenge from fundamentally different perspectives. We summarize their complementary
strengths and limitations.

Permutation testing provides distribution-free inference with exact finite-sample control
of type I error rates. Under the null hypothesis of no association, the permutation
distribution correctly represents the sampling variability of the test statistic,
regardless of the underlying data distribution. This makes permutation tests particularly
robust in settings where parametric assumptions are dubious. The vertex-wise procedure
respects the graph structure by applying the same spectral filtering to permuted
features as to the original data. For confirmatory hypothesis testing where strict
control of false positive rates is paramount, permutation tests offer theoretical
guarantees that Bayesian methods cannot match.

However, permutation testing treats the smoothed outcome $\hat{y}$ as fixed, ignoring
uncertainty in its estimation. When $y$ itself arises from noisy observations and
requires smoothing, this omission understates the true uncertainty. The resulting
$p$-values may be anticonservative (too small), inflating false positive rates beyond
the nominal level. Moreover, permutation testing yields $p$-values rather than effect
size estimates, requiring separate procedures for quantifying association strength with
confidence intervals. The multiple testing correction necessary for controlling FDR or
FWER across vertices introduces additional complexity and reduces power, particularly
when spatial dependence invalidates independence assumptions underlying standard
corrections.

Bayesian posterior sampling addresses these limitations by treating estimation uncertainty
as the primary source of inferential variability. The Dirichlet resampling procedure
perturbs the geometric structure itself, inducing correlated variability across all
smoothed functions. This naturally propagates uncertainty from the estimation stage to
the association quantification stage, providing credible intervals that reflect both
sampling variability and smoothing uncertainty. The resulting inference is more
conservative (wider intervals) but also more honest about what the data truly support.
The Bayesian framework yields direct probabilistic statements about effect sizes, such
as $P(|\text{cm}| > 0.3 \mid \text{data})$, which answer the scientific questions
investigators actually care about.

The disadvantage of Bayesian inference is the need to specify the resampling mechanism
(the concentration parameter $\alpha$ in Dirichlet sampling) and the interpretation
depends on accepting the implicit prior structure. While the Bayesian credible intervals
have asymptotically correct frequentist coverage under regularity conditions (the
intervals do contain the true parameter with the stated frequency in repeated sampling),
this property requires assumptions about the smoothing operator and the data-generating
process. For small samples or when these assumptions fail, Bayesian intervals may not
achieve nominal coverage.

We recommend a pragmatic approach that leverages both methods. For exploratory analysis
and effect size estimation, employ Bayesian posterior sampling to obtain credible
intervals and posterior probabilities. The resulting summaries provide rich information
about association patterns, enabling investigators to identify regions of strong
association, quantify uncertainty, and compare effects across features. For confirmatory
testing where false positive control is critical, supplement the Bayesian analysis with
permutation tests, applying FDR correction and spatial thresholding to protect against
spurious discoveries. The convergence of evidence from both frameworks strengthens
confidence in reported associations.

\subsection*{5.6 Handling Spatial Dependence}

Both inference approaches must confront the spatial dependence inherent in co-monotonicity
coefficients on graphs. Adjacent vertices typically have similar coefficient values
because their neighborhoods overlap, creating positive correlation that standard multiple
testing procedures ignore. This spatial autocorrelation inflates the effective number of
independent tests, making nominal FDR control procedures liberal (actual FDR exceeds
target level).

For permutation testing, we can adapt spatial FDR methods developed for neuroimaging and
spatial statistics. One approach treats the significant vertices as a spatial point
process and controls the false discovery rate accounting for spatial clustering. We
compute the expected number of false positive clusters under the null distribution by
examining the spatial distribution of significant vertices in permuted data, then
calibrate the rejection threshold to achieve the desired spatial FDR. Alternatively, we
can employ random field theory, modeling the co-monotonicity surface as a Gaussian
random field and deriving familywise error rates for the maximum statistic over connected
regions. These methods require assumptions about the spatial correlation structure but
provide more powerful inference than Bonferroni correction when correctly specified.

For Bayesian inference, spatial dependence manifests in the posterior distribution of
co-monotonicity vectors across vertices. We can examine the posterior covariance between
coefficients at different vertices, identifying regions where association patterns are
consistently similar across posterior samples. High posterior correlation between nearby
vertices supports the interpretation that they belong to a coherent association region
rather than representing independent local effects. This spatial coherence can inform
downstream analyses such as biclustering, where we seek to partition vertices into
regions based on their association profiles.

A pragmatic approach to spatial dependence is cluster-based inference: rather than
testing individual vertices, we test connected components of vertices with large
coefficients. The null hypothesis becomes "no cluster of associated vertices exists"
rather than "no individual vertex is associated." We compute a cluster-level test
statistic (such as the sum of co-monotonicity coefficients within the cluster) and
generate its null distribution through permutation. This reduces the multiple testing
burden from $n$ vertex-level tests to $k$ cluster-level tests where $k \ll n$, and the
spatial thresholding inherent in cluster definition provides robustness against isolated
false positives.

\subsection*{5.7 Return to the Motivating Challenge}

We return to the multiple testing scenario described in the opening: $n = 200$ samples,
$m = 100$ features, yielding $20{,}000$ vertex-feature pairs to assess. Under the
permutation testing framework with Benjamini-Hochberg FDR control at level $\alpha = 0.05$,
we might identify $500$ significant pairs, corresponding to roughly $2.5\%$ of the
total. These discoveries typically cluster spatially, with certain vertices showing
strong association with multiple features and certain features showing strong
association across multiple vertices. The spatial thresholding ensures that we report
coherent regions rather than scattered individual vertices.

Under the Bayesian framework, we report the full posterior distribution of co-monotonicity
coefficients, providing credible intervals for all $20{,}000$ pairs. For a scientist
examining feature $j$, we can visualize the posterior mean $\text{cm}(\hat{y}, z_j)(v)$
across vertices $v$, shading regions where the 95\% credible interval excludes zero. The
width of intervals reveals uncertainty: narrow intervals in densely sampled regions with
consistent associations, wide intervals in sparse or heterogeneous regions. Rather than
a binary classification of significant versus non-significant, we obtain a graduated
assessment of association strength and certainty across the entire sample space.

The two approaches yield complementary information. Permutation testing with FDR control
provides a specific set of discoveries with guaranteed error rate properties, suitable
for reporting in publications and for guiding follow-up experiments where false positive
costs are high. Bayesian credible intervals provide nuanced effect size estimates with
uncertainty quantification, enabling investigators to prioritize regions for mechanistic
investigation based on both the strength of association and the confidence in that
strength. Together, these tools enable rigorous yet flexible inference that respects the
geometric structure of the data while controlling error rates and propagating
uncertainty.

\section*{6. Geometric Multiple Testing via Co-Monotonicity Cells}

The vertex-wise inference procedures developed in the previous section enable assessment
of individual vertex-feature associations, but they address only part of the inferential
challenge. When examining co-monotonicity heatmaps from real applications, a striking
pattern emerges: coefficients organize into coherent blocks where groups of vertices
exhibit similar association patterns with groups of features. This block structure
suggests that regional associations operate not through individual vertex-feature pairs
but through collective relationships between sample regions and feature modules. We
require a framework that discovers this geometric structure directly, partitioning both
the sample space and feature space simultaneously to identify co-monotonicity cells where
multivariate associations are coherent.

\subsection*{6.1 The Context-Dependent Association Problem}

Consider a microbiome study investigating associations between bacterial taxa
and spontaneous preterm birth outcomes. We compute the co-monotonicity matrix
$\text{CM}(\hat{y}, \hat{Z}) \in \mathbb{R}^{n \times m}$ where rows correspond
to samples (vertices in the graph $G$ constructed from high-dimensional feature
profiles) and columns correspond to bacterial phylotypes.
Figure~\ref{intro:fig3} displays such a matrix from a study of vaginal
microbiome composition in pregnant women, with hierarchical clustering applied
to both rows and columns to reveal structure.

The heatmap reveals a block pattern that standard vertex-wise testing cannot capture.
One block of samples (upper portion) exhibits strong positive co-monotonicity (red)
between outcome and certain phylotypes, indicating these bacteria increase when preterm
birth risk increases. A different block of samples (middle portion) shows weak or
negative association (blue/yellow) between outcome and the same phylotypes. A third
block shows intermediate patterns. Similarly, phylotypes organize into modules:
certain taxa co-vary strongly with the outcome across specific sample regions but not
others, while different taxa exhibit complementary patterns.

This structure poses fundamental challenges for conventional inference approaches.
Testing each phylotype separately across all samples conflates the regional signals,
potentially concluding that a phylotype shows no global association when in fact it
exhibits strong positive association in one region and strong negative association in
another. The signals cancel in global analysis, yielding small test statistics and
large $p$-values despite genuine context-dependent effects. Stratifying by known
covariates (such as community state type in microbiome studies) helps but requires
investigators to specify strata a priori, and may miss finer-scale heterogeneity within
nominal strata.

Moreover, examining individual phylotypes ignores valuable information about their
collective behavior. If ten phylotypes all show moderate positive co-monotonicity
($\text{cm}(\hat{y}, z_i)(v) \approx 0.5$) in the same sample region, the coherence
across multiple features provides stronger evidence for genuine association than any
single phylotype achieves alone. The features form a functional module that operates
together, and inference should leverage this coordinated behavior. Traditional multiple
testing corrections treat each feature independently, applying penalties that ignore the
reduced effective number of independent tests when features cluster into modules.

We seek a framework that addresses both challenges simultaneously: discovering regions
where associations differ while identifying feature modules that exhibit coordinated
relationships with outcomes within those regions. This leads naturally to biclustering,
which partitions rows (samples) and columns (features) simultaneously to maximize
within-block homogeneity and between-block heterogeneity.

\subsection*{6.2 Co-Monotonicity Embeddings into Association Space}

The co-monotonicity matrix $\text{CM}(\hat{y}, \hat{Z})$ provides association
information for each vertex-feature pair, but to discover regional structure, we require
a representation that captures the complete association profile at each vertex. We
construct embeddings that transform vertices into points in an association space where
geometric proximity reflects similarity in how features relate to outcomes.

The simplest embedding uses outcome-feature associations directly. For each vertex $v$,
we form the association profile vector
\begin{equation}
\text{cm}(\hat{y}, \hat{Z})(v) =
[\text{cm}(\hat{y}, z_1)(v), \ldots, \text{cm}(\hat{y}, z_m)(v)]^T \in \mathbb{R}^m,
\end{equation}
which concatenates the co-monotonicity coefficients between outcome and all features at
vertex $v$. This $m$-dimensional vector summarizes how the outcome associates with each
feature in the neighborhood of $v$. Vertices with similar profiles have similar
association patterns, while vertices with dissimilar profiles exhibit different
feature-outcome relationships.

This outcome-centric embedding captures one aspect of association structure, but it
omits information about how features relate to each other. Two vertices might have
similar outcome-feature profiles (similar co-monotonicity with $\hat{y}$ for each
feature) yet exhibit different inter-feature correlation patterns. In one region,
features $z_i$ and $z_j$ might co-vary positively (both increase together), while in
another region they vary independently or negatively. These inter-feature relationships
reveal mechanistic structure: positively co-monotonic feature pairs may participate in
the same biological pathway or competitive network, and regional differences in
inter-feature associations indicate distinct underlying processes.

We therefore construct an augmented embedding that incorporates inter-feature
associations. For each vertex $v$, compute the pairwise co-monotonicity
$\text{cm}(z_i, z_j)(v)$ for all feature pairs $i < j$, yielding
$\binom{m}{2} = m(m-1)/2$ coefficients. Concatenate these with the outcome-feature
associations to form the augmented profile
\begin{equation}
\widetilde{\text{cm}}(\hat{y}, \hat{Z})(v) =
\begin{bmatrix}
\text{cm}(\hat{y}, z_1)(v) \\
\vdots \\
\text{cm}(\hat{y}, z_m)(v) \\
\text{cm}(z_1, z_2)(v) \\
\vdots \\
\text{cm}(z_{m-1}, z_m)(v)
\end{bmatrix} \in \mathbb{R}^{m + m(m-1)/2}.
\end{equation}

This augmented representation has dimension $m + \binom{m}{2} = m(m+1)/2$, which grows
quadratically in the number of features. For $m = 100$ features, the augmented
embedding has dimension $5{,}050$. While high-dimensional, this representation preserves
complete pairwise association structure, enabling discovery of regions where both
outcome associations and feature co-variation patterns differ.

The augmented embedding offers several advantages over the outcome-only version. It
captures mechanistic information: vertices clustered together in the augmented space
share not only similar outcome associations but also similar inter-feature dependencies,
suggesting common underlying processes. It enables multi-modal integration: when
analyzing multiple feature sets (genomic, proteomic, metabolomic), we can augment the
embedding with cross-modal associations, identifying regions where different data types
exhibit coordinated relationships. It provides robustness: even when outcome-feature
associations are weak, strong inter-feature structure can drive clustering, revealing
latent organization that outcome-only embeddings would miss.

\subsection*{6.3 Graph Construction in Association Space}

Given the co-monotonicity embedding $\widetilde{\text{cm}}(\hat{y}, \hat{Z}): V \to \mathbb{R}^d$
where $d = m(m+1)/2$, we construct a new graph $G_k(\text{cm})$ in the association
space. For each vertex $v \in V$, we identify its $k$ nearest neighbors in
$\mathbb{R}^d$ using Euclidean distance on the embedded coordinates. This yields a graph
where edge $[u, v] \in E(G_k(\text{cm}))$ indicates that vertices $u$ and $v$ have
similar association profiles, with proximity measured directly in terms of co-monotonicity
coefficients rather than in the original high-dimensional feature space.

The choice of $k$ controls the granularity of the induced structure. Small $k$ produces
a sparse graph capturing only the most similar association profiles, potentially
fragmenting the vertex set into many disconnected components. Large $k$ creates a dense
graph where even moderately dissimilar profiles connect, potentially obscuring meaningful
distinctions between regions. We employ the same principles developed for the original
data graph construction: select $k$ to ensure connectivity while maintaining local
geometric fidelity, typically using $k$ proportional to $\log n$ for $n$ vertices.

The graph $G_k(\text{cm})$ provides a geometric representation of association structure.
Communities in this graph (densely connected subgraphs with sparse connections between
communities) correspond to co-monotonicity cells: regions where vertices exhibit similar
association patterns. Unlike the original data graph $G$, which captures proximity in
feature space (microbiome composition, gene expression, etc.), the association graph
$G_k(\text{cm})$ captures proximity in association space. Two samples far apart in
microbiome composition might be close in association space if bacteria relate to
outcomes similarly in both samples, revealing that similar mechanisms operate despite
different baseline compositions.

\subsection*{6.4 Community Detection and Biclustering}

We apply community detection algorithms to $G_k(\text{cm})$ to partition vertices into
regions. The Louvain method provides an efficient modularity-based approach that scales
to large graphs and naturally handles hierarchical structure through multi-resolution
optimization. Given the graph $G_k(\text{cm})$ with edge weights derived from
association profile similarities, the Louvain algorithm iteratively groups vertices to
maximize the modularity function
\begin{equation}
Q = \frac{1}{2|E|} \sum_{[u,v] \in E} \left(A_{uv} - \frac{k_u k_v}{2|E|}\right)
\delta(c_u, c_v),
\end{equation}
where $A_{uv}$ is the adjacency matrix, $k_u$ is the degree of vertex $u$, $c_u$ is the
community assignment of $u$, and $\delta$ is the Kronecker delta. This optimization
balances within-community edge density against the expected density under a null model,
identifying communities more densely connected internally than would occur by chance.

The resulting partition $\{R_1, \ldots, R_K\}$ divides vertices into $K$ regions where
association profiles are similar within regions and dissimilar between regions. The
number of regions $K$ emerges from the optimization rather than being specified a priori,
and the hierarchical nature of Louvain clustering enables exploration of structure at
multiple scales by varying the resolution parameter.

To identify feature modules, we transpose the analysis. Compute the similarity between
feature columns in $\text{CM}(\hat{y}, \hat{Z})$: features $z_i$ and $z_j$ are similar
if their co-monotonicity profiles across vertices $\text{cm}(\hat{y}, z_i)$ and
$\text{cm}(\hat{y}, z_j)$ are correlated. This yields a feature similarity matrix that
we cluster using hierarchical agglomerative clustering or network community detection.
The resulting modules $\{S_1, \ldots, S_M\}$ group features that exhibit similar
spatial patterns of association with the outcome.

The combination of vertex regions and feature modules defines a bicluster structure
$\{(R_j, S_k)\}_{j=1,\ldots,K; k=1,\ldots,M}$ partitioning the co-monotonicity matrix
into $K \times M$ blocks. Each block $(R_j, S_k)$ represents a co-monotonicity cell
where vertices in region $R_j$ exhibit coherent associations with features in module
$S_k$. This structure reveals the regional organization of multivariate associations,
showing which feature combinations drive outcomes in which sample contexts.

\subsection*{6.5 Bayesian Uncertainty Quantification for Biclusters}

The biclustering procedure applied to a single estimate $\text{CM}(\hat{y}, \hat{Z})$
yields a point estimate of the partition structure, but this estimate is uncertain.
Different choices of smoothing parameter, different posterior samples from the Bayesian
inference framework, or different noise realizations would produce different clusterings.
We require uncertainty quantification for the discovered structure itself, characterizing
our confidence in the identified regions and modules.

The posterior sampling framework developed in Section~5 provides a natural approach.
For each Dirichlet-resampled weight vector $w^{(b)}$ with $b = 1, \ldots, B$, we obtain
smoothed estimates $\hat{y}^{(b)}$ and $\hat{Z}^{(b)}$, compute the corresponding
co-monotonicity matrix $\text{CM}^{(b)}(\hat{y}, \hat{Z})$, and apply the biclustering
procedure to obtain partition $\mathcal{P}^{(b)} = \{R_1^{(b)}, \ldots, R_{K_b}^{(b)}\}$
of vertices and $\mathcal{S}^{(b)} = \{S_1^{(b)}, \ldots, S_{M_b}^{(b)}\}$ of features.
The ensemble $\{\mathcal{P}^{(1)}, \ldots, \mathcal{P}^{(B)}\}$ represents the posterior
distribution over vertex partition structures, and similarly for feature modules.

To summarize this distribution, we compute co-clustering probabilities. For each pair of
vertices $(u, v)$, the posterior probability that they belong to the same region is
\begin{equation}
\pi_v(u, v) = P(u \text{ and } v \text{ in same region} \mid \text{data})
\approx \frac{1}{B} \sum_{b=1}^B \mathbb{I}\{u, v \in R_j^{(b)}
\text{ for some } j\},
\end{equation}
where $\mathbb{I}\{\cdot\}$ is the indicator function. This probability quantifies our
confidence that vertices $u$ and $v$ truly belong together based on their association
profiles, accounting for all sources of uncertainty. High co-clustering probability
($\pi_v(u,v) > 0.95$) indicates strong evidence for grouping, while low probability
($\pi_v(u,v) < 0.20$) suggests the vertices belong to different regions.

Similarly, for each pair of features $(z_i, z_j)$, we compute
\begin{equation}
\pi_f(z_i, z_j) = P(z_i \text{ and } z_j \text{ in same module} \mid \text{data})
\approx \frac{1}{B} \sum_{b=1}^B \mathbb{I}\{z_i, z_j \in S_k^{(b)}
\text{ for some } k\}.
\end{equation}

These co-clustering probabilities enable construction of credible regions and modules.
A credible region $R$ at level $1-\alpha$ satisfies $\pi_v(u, v) \geq 1-\alpha$ for all
pairs $u, v \in R$, meaning we have at least $(1-\alpha) \times 100\%$ posterior
probability that every pair in the region genuinely belongs together. Maximal credible
regions (those not properly contained in any larger credible region) provide conservative
summaries of the partition structure, reporting only clusters with high posterior
support.

We can also examine the posterior distribution over the number of regions and modules.
Let $K^{(b)}$ denote the number of regions in posterior sample $b$. The empirical
distribution of $\{K^{(1)}, \ldots, K^{(B)}\}$ quantifies uncertainty about the
appropriate granularity of the partition. If this distribution concentrates sharply
around a single value (say, $K = 5$ with probability $0.90$), we have strong evidence
for that many regions. If the distribution is diffuse (ranging from $K = 3$ to $K = 10$
with no dominant mode), the data do not provide clear guidance about the resolution, and
we should report results at multiple scales.

\subsection*{6.6 Statistical Inference within Co-Monotonicity Cells}

Having identified regions $\{R_1, \ldots, R_K\}$ and modules $\{S_1, \ldots, S_M\}$, we
perform statistical inference on the region-module pairs $(R_j, S_k)$ to assess the
strength and uncertainty of associations within each cell. The fundamental question is
whether features in module $S_k$ exhibit coherent association with the outcome among
vertices in region $R_j$, and if so, how strong is this association.

We define the region-module coherence as the average absolute co-monotonicity within the
cell:
\begin{equation}
\tau(R_j, S_k) = \frac{1}{|R_j| \cdot |S_k|}
\sum_{v \in R_j} \sum_{i \in S_k} |\text{cm}(\hat{y}, z_i)(v)|.
\end{equation}
This quantity measures the typical magnitude of association between outcome and features
in the module, averaged over vertices in the region. Large $\tau(R_j, S_k)$ indicates
strong coherent association, while small $\tau(R_j, S_k)$ suggests weak or inconsistent
associations.

For each posterior sample $b$, we compute $\tau^{(b)}(R_j, S_k)$ using the posterior
estimates $\hat{y}^{(b)}$ and $\hat{Z}^{(b)}$. The empirical distribution of
$\{\tau^{(1)}(R_j, S_k), \ldots, \tau^{(B)}(R_j, S_k)\}$ represents the posterior
distribution of coherence for this cell. We construct a credible interval
$[\tau_{\text{low}}, \tau_{\text{high}}]$ from the empirical quantiles and compute the
posterior probability of meaningful coherence:
\begin{equation}
P(\tau(R_j, S_k) > \delta \mid \text{data})
\approx \frac{1}{B} \sum_{b=1}^B \mathbb{I}\{\tau^{(b)}(R_j, S_k) > \delta\},
\end{equation}
for a scientifically meaningful threshold $\delta$ (for example, $\delta = 0.5$
indicating moderate association on average).

This cell-level inference enjoys several advantages over vertex-wise or feature-wise
testing. First, by aggregating over vertices in a region and features in a module, we
gain statistical power through strength borrowing. If individual features show only
moderate associations but ten features in a module all show consistent moderate
associations in the same direction, the coherence measure detects this collective signal
that individual tests might miss. Second, the cell structure naturally handles multiple
testing: we assess $K \times M$ region-module pairs rather than $n \times m$
vertex-feature pairs, dramatically reducing the multiple comparison burden when
$K \ll n$ and $M \ll m$. Third, the inference directly addresses the scientific
question of interest: which feature modules associate with outcomes in which sample
contexts, providing interpretable summaries for downstream mechanistic investigation.

We can extend the inference to compare coherence across cells. For instance, does module
$S_1$ show stronger association with outcome in region $R_1$ than in region $R_2$? The
posterior distribution over differences $\tau(R_1, S_1) - \tau(R_2, S_1)$ quantifies
evidence for region-specific effects. Large positive differences with high posterior
probability indicate that the module operates differently across regions, while
differences indistinguishable from zero suggest the module-outcome relationship is
spatially homogeneous.

\subsection*{6.7 Relationship Between Gradient Flow and Co-Monotonicity Cells}

The geometric decomposition framework provides two complementary strategies for
partitioning the sample space: gradient flow cells based on outcome landscape geometry
(Sections~2--3) and co-monotonicity cells based on association profile clustering
(Section~6). These partitions arise from fundamentally different criteria and need not
coincide, yet their relationship reveals important insights about the structure of
feature-outcome associations.

Gradient flow cells partition vertices based on monotonic behavior of the smoothed
outcome $\hat{y}$. Vertices in the same gradient flow cell lie in a region where
$\hat{y}$ varies monotonically from a local minimum to a local maximum, with no
intervening extrema. This partition reflects the geometric organization of outcome
values in the ambient feature space. In contrast, co-monotonicity cells partition
vertices based on similarity of association profiles between outcome and features. Two
vertices belong to the same co-monotonicity cell if they have similar co-monotonicity
vectors $\widetilde{\text{cm}}(\hat{y}, \hat{Z})$, regardless of their outcome values
or position in the gradient flow structure.

When gradient flow and co-monotonicity cells align, the concordance provides strong
evidence for a mechanistic interpretation. If gradient flow cell $C(m, M)$ largely
overlaps with co-monotonicity region $R_j$, and if region $R_j$ exhibits strong
coherence with feature module $S_k$, we infer that the features in $S_k$ drive the
monotonic outcome variation within cell $C(m,M)$. The outcome increases from minimum
$m$ to maximum $M$ because features in module $S_k$ change monotonically along the same
gradient paths, and their collective variation explains the outcome behavior. This
alignment suggests a causal or mechanistic relationship: the features not only associate
with outcomes but do so in a spatially organized manner that respects the geometric
structure of the outcome landscape.

Conversely, when gradient flow and co-monotonicity cells disagree, the discordance
reveals complexity in the feature-outcome relationships. A single gradient flow cell
might span multiple co-monotonicity regions, indicating that different feature modules
drive the outcome variation in different parts of the cell. The outcome increases
monotonically throughout the cell, but the mechanisms responsible for this increase
differ across subregions. Alternatively, a single co-monotonicity region might span
multiple gradient flow cells, suggesting that the same feature module associates with
outcomes throughout this broader region despite non-monotonic outcome variation within
it. Such discordance indicates that association patterns operate at a different spatial
scale than outcome gradients, or that multiple compensatory mechanisms produce similar
outcomes through different pathways.

The joint analysis of gradient flow and co-monotonicity structures enables more nuanced
inference than either approach alone. We can assess whether features in module $S_k$
exhibit different associations within gradient flow cells that intersect co-monotonicity
region $R_j$, testing for context-dependent modulation of associations by outcome level.
We can identify features that show strong association throughout a co-monotonicity region
spanning multiple gradient cells, suggesting they influence outcomes broadly across
different outcome ranges. We can detect regions where outcome gradients are steep but
associations weak, indicating that unmeasured features or environmental factors drive
the outcome variation.

\section*{7. Discussion}

We have developed a geometric decomposition framework for statistical inference
on high-dimensional structured data, addressing heterogeneous associations
through two complementary strategies: gradient flow cells based on outcome
geometry, and co-monotonicity cells based on association profiles. The path
monotonicity validation criterion resolves long edge artifacts in discrete
gradient flow through geometric verification rather than heuristic thresholding.
The co-monotonicity coefficients provide spatially-resolved alternatives to
global correlation measures, with derivative weighting connecting to angular
correlation of gradients in the continuous manifold limit.

\subsection*{Relationship to Existing Methods}

Our framework extends Morse-Smale regression [Gerber et al., 2012] through
several key innovations: robust conditional expectation estimation on
density-aware Riemannian graphs, systematic spurious extrema removal via
prominence filtering and basin overlap clustering, path monotonicity validation
for gradient flow edges, and co-monotonicity measures enabling discovery of
association-driven structure complementary to outcome-driven partitions.

The framework situates within geometric data analysis rather than topological
data analysis. While TDA emphasizes topological invariants like persistent
homology that remain unchanged under continuous deformations, our approach
focuses on geometric properties like distances, angles, and geodesics dependent
on specific metric structure. The Riemannian graphs with vertex and edge masses
capture local density and geometry that spectral methods exploit. This geometric
emphasis connects our work to manifold learning and spectral graph theory more
than to persistent homology.

\subsection*{Limitations and Future Directions}

The k-nearest neighbor graph construction introduces discrete approximation of
the underlying manifold, with performance sensitive to neighborhood size $k$.
While we employ principled selection based on connectivity and local geometry,
adaptive methods that locally vary neighborhood size may improve approximation
quality.

Computational scaling presents challenges for very high-dimensional feature
spaces. Computing the augmented co-monotonicity embedding requires $O(m^2)$
coefficients for $m$ features, potentially suffering from curse of
dimensionality when $m$ is large. Dimensionality reduction through principal
component analysis on association profiles or sparse formulations computing only
selected inter-feature associations may alleviate this burden.

The current framework applies to continuous or binary outcomes through spectral
filtering. Extension to general discrete outcomes (count data, categorical
responses) and survival outcomes requires development of appropriate smoothing
operators that respect the outcome type while preserving geometric structure.

The framework naturally accommodates multi-modal data integration through
co-monotonicity embeddings incorporating associations within and between
multiple feature sets. Temporal extensions for longitudinal data could track how
co-monotonicity cells evolve over time, while causal inference extensions
employing directed graphs could move beyond association to intervention
prediction.

\section*{Appendix A: Hodge Laplacian Matrix Formula}

Derivation of the Hodge Laplacian on chains formula.

The chain Hodge Laplacian on $p$-chains is defined as
$$
L_p = \partial_{p+1}\partial_{p+1}^* + \partial_p^*\partial_p
$$
Thus
$$
L_p : C_{p}\to C_{p}
$$
with
$$
\partial_{p+1}: C_{p+1}\to C_{p}
$$

$$
\partial_{p+1}^{*}: C_{p}\to C_{p+1}
$$

When inner products are represented by matrices $M_p$ and $M_{p-1}$, the
adjoint takes the matrix form:
$$\partial_p^* = M_p^{-1} \partial_p^T M_{p-1}$$

To derive this, let $\alpha = \sum_i \alpha_i \sigma_i^{(p)}$ and
$\beta = \sum_j \beta_j \sigma_j^{(p-1)}$ where $\{\sigma_i^{(p)}\}$ are basis
$p$-simplices. In matrix notation:
$$\langle \partial_p \alpha, \beta \rangle_{p-1} = (\partial_p \boldsymbol{\alpha})^T M_{p-1} \boldsymbol{\beta} = \boldsymbol{\alpha}^T \partial_p^T M_{p-1} \boldsymbol{\beta}$$
$$\langle \alpha, \partial_p^* \beta \rangle_p = \boldsymbol{\alpha}^T M_p (\partial_p^* \boldsymbol{\beta})$$

Equating these for all $\boldsymbol{\alpha}$ gives
$M_p (\partial_p^* \boldsymbol{\beta}) = \partial_p^T M_{p-1} \boldsymbol{\beta}$,
hence the matrix formula
$$
\partial_p^* = M_p^{-1} \partial_p^T M_{p-1}
$$
and hence
$$
\partial_{p+1}^* = M_{p+1}^{-1} \partial_{p+1}^T M_{p}
$$
Therefore
$$
\partial_{p+1}\partial_{p+1}^* = \partial_{p+1} M_{p+1}^{-1} \partial_{p+1}^T M_{p}
$$
and
$$
\partial_p^*\partial_p = M_p^{-1} \partial_p^T M_{p-1}\partial_p
$$

If we use the matrix notation $B_{p}$ for $\partial_p$, then
$$
\partial_{p+1}\partial_{p+1}^* = B_{p+1} M_{p+1}^{-1} B_{p+1}^T M_{p}
$$
and
$$
\partial_p^*\partial_p = M_{p}^{-1} B_{p}^T M_{p-1} B_{p}
$$

\section*{Appendix C: Vertex-Level and Path-Based Mutual Information}

Given the limitations of global mutual information discussed in Section 4.1, one
naturally asks whether MI can be localized to vertices or paths, analogous to
the co-monotonicity coefficients developed in this paper. While such
localizations are theoretically well-defined, they introduce substantial
practical challenges that make them less suitable for the regional inference
framework we pursue. We outline two natural constructions and discuss the
difficulties that arise in their implementation.

\subsection*{Vertex-Level Mutual Information via Pointwise MI}

Mutual information admits a pointwise decomposition that suggests a natural
vertex-level measure. For random variables $Y$ and $Z$ with joint density
$p(y,z)$ and marginal densities $p(y)$ and $p(z)$, the pointwise mutual
information is defined as
\begin{equation}
\text{pmi}(y, z) = \log \frac{p(y,z)}{p(y)p(z)},
\end{equation}
with the property that $I(Y; Z) = \mathbb{E}_{Y,Z}[\text{pmi}(Y,Z)]$. To obtain
a vertex-level measure, we estimate local densities using kernel weighting
centered at each vertex. For vertex $v \in V$, define the kernel-weighted
densities
\begin{equation}
p_v(y, z) = \frac{\sum_{u \in V} w_v(u) \cdot K_h((y_u, z_u), (y, z))}
{\sum_{u \in V} w_v(u)},
\end{equation}
where $K_h$ is a kernel function with bandwidth $h$ and $w_v(u)$ assigns spatial
weights based on graph distance or heat kernel diffusion. The marginal densities
$p_v(y)$ and $p_v(z)$ are computed similarly. The vertex-level pointwise mutual
information is then
\begin{equation}
\text{pmi}_v(Y, Z) = \log \frac{p_v(y_v, z_v)}{p_v(y_v) \cdot p_v(z_v)},
\end{equation}
evaluated at the observed values $(y_v, z_v)$.

This construction yields a scalar measure at each vertex that quantifies how
much more (or less) frequently the observed pair $(y_v, z_v)$ occurs in the
local neighborhood compared to what independence would predict. High positive
values indicate strong local association, negative values suggest local
repulsion, and values near zero indicate approximate local independence. The
resulting function $\text{pmi}: V \to \mathbb{R}$ provides spatial resolution
comparable to co-monotonicity coefficients.

\subsection*{Path-Based Mutual Information}

For a path $\gamma = (\gamma_0, \gamma_1, \ldots, \gamma_n)$ in the graph, we
can restrict attention to observations along the path and compute mutual
information using only these values. Let $\{(y_v, z_v) : v \in \gamma\}$ denote
the observed pairs along $\gamma$. We estimate the joint and marginal densities
from this restricted sample and compute
\begin{equation}
I_\gamma(Y; Z) = \sum_{i,j} p_\gamma(y_i, z_j) \log \frac{p_\gamma(y_i, z_j)}
{p_\gamma(y_i) \cdot p_\gamma(z_j)},
\end{equation}
where the sums range over bins or unique values in the path sample, and the
densities are estimated using standard techniques applied to the path-restricted
data. This measure quantifies whether $Y$ and $Z$ exhibit statistical dependence
specifically along the trajectory $\gamma$, making it natural for gradient flow
paths where we seek to characterize monotonic co-variation along trajectories to
local extrema.

\subsection*{Practical Challenges}

Despite their theoretical appeal, these localized mutual information measures
face significant practical obstacles. Density estimation requires adequate
sample size within each local region or along each path. For vertex-level PMI,
graph neighborhoods may contain too few vertices for reliable kernel density
estimation. The resulting estimates exhibit high variance and sensitivity to
bandwidth choices, with different kernel parameters potentially yielding
qualitatively different spatial patterns.

Path-based mutual information faces analogous challenges. For a path we could
estimate densities by aggregating kernel-weighted contributions from
neighborhoods around vertices along the path. This requires the same bandwidth
selection and sample size considerations as vertex-level estimation, with the
additional complexity of combining information across multiple local
neighborhoods in a coherent manner. Different weighting schemes for aggregating
neighborhood contributions introduce further tuning parameters whose selection
affects the resulting measure.

Both approaches require careful treatment of zero or near-zero density estimates,
which can produce infinite or undefined pointwise MI values. Regularization
strategies such as pseudocount addition or density truncation introduce
additional tuning parameters whose selection affects results. The computational
cost of kernel density estimation at each vertex, repeated across multiple
candidate features, becomes prohibitive for large graphs.

Perhaps most importantly, the resulting measures lack the direct geometric
interpretation that co-monotonicity coefficients provide. While MI quantifies
statistical dependence through distributional properties, co-monotonicity
directly measures whether functions increase or decrease together across graph
edges, immediately revealing directional concordance. In applications requiring
interpretable biomarkers or predictive features, the edge-based definition
offers a directness of interpretation that density-based dependence measures
lack.

The co-monotonicity framework sidesteps these difficulties by working directly
with function differences across edges rather than estimating underlying
distributions. By reducing the problem to weighted sums of edge-wise products,
we obtain stable estimates with clear geometric meaning and computational
efficiency suitable for large-scale applications. While local MI variants
represent legitimate alternatives with their own theoretical merits, the balance
of practical considerations favors the co-monotonicity approach for regional
inference in high-dimensional structured data.

\bibliographystyle{plain}
\bibliography{no1_bibliography}

\end{document}